\documentclass[prb,twocolumn,amsmath,amssymb,floatfix,eqsecnum]{revtex4}
\usepackage{graphicx}
\usepackage{bm}
\usepackage{hyperref}

\begin{document}

\preprint{cond-mat/0205169}

\title{Mott insulators in strong electric fields}

\author{Subir Sachdev}
\email{subir.sachdev@yale.edu}
\homepage{http://pantheon.yale.edu/~subir} \affiliation{Department
of Physics, Yale University, P.O. Box 208120, New Haven CT
06520-8120}
\author{K. Sengupta}
\email{k.sengupta@yale.edu} \affiliation{Department of Physics,
Yale University, P.O. Box 208120, New Haven CT 06520-8120}
\author{S. M. Girvin}
\email{steven.girvin@yale.edu}
\homepage{http://pantheon.yale.edu/~smg47} \affiliation{Department
of Physics, Yale University, P.O. Box 208120, New Haven CT
06520-8120}

\date{May 8, 2002}

\begin{abstract}
Recent experiments on ultracold atomic gases in an optical lattice
potential have produced a Mott insulating state of $^{87}$Rb
atoms. This state is stable to a small applied potential gradient
(an `electric' field), but a resonant response was observed when
the potential energy drop per lattice spacing ($E$), was close to
the repulsive interaction energy ($U$) between two atoms in the
same lattice potential well. We identify all states which are
resonantly coupled to the Mott insulator for $E \approx U$ via an
infinitesimal tunneling amplitude between neighboring potential
wells. The strong correlation between these states is described by
an effective Hamiltonian for the resonant subspace. This
Hamiltonian exhibits quantum phase transitions associated with an
Ising density wave order, and with the appearance of superfluidity
in the directions transverse to the electric field. We suggest
that the observed resonant response is related to these
transitions, and propose experiments to directly detect the order
parameters. The generalizations to electric fields applied in
different directions, and to a variety of lattices, should allow
study of numerous other correlated quantum phases.
\end{abstract}

\maketitle

\section{Introduction}
\label{intro} Recent experiments on ultracold trapped atomic gases
have opened a new window onto the phases of quantum matter
\cite{mark,bloch}. A gas of bosonic atoms has been reversibly
tuned between superfluid and insulating ground states by varying
the strength of a periodic potential produced by standing waves of
laser light\cite{bloch}. These experiments offer unprecedented
control of the microscopic parameters, and allow exploration of
parameter regimes not previously available in analogous condensed
matter systems.

\begin{figure}
\centerline{\includegraphics[width=2.2in]{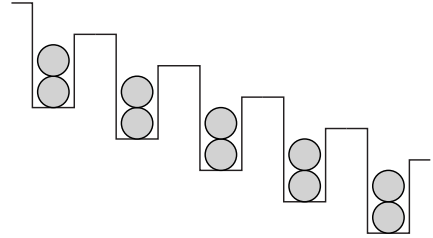}}
\caption{Figures~\protect\ref{fig1a}-\protect\ref{fig1eg} contain
schematic representations of the Mott insulator, and of various
states coupled to it. Shown above is the Mott insulator with
$n_0=2$. Each well represents a local minimum of the optical
lattice potential - these we number 1-5 from the left. The
potential gradient leads to a uniform decrease in the on-site
energy of atom as we move to the right. The grey circles are the
$b_i$ bosons of (\protect\ref{hubbard}). The vertical direction
represents increasing energy: the repulsive interaction energy
between the atoms is realized by placing atoms vertically within
each well, so that each atom displaces the remaining atoms upwards
along the energy axis. We have chosen the diameter of the atoms to
equal the potential energy drop between neighboring wells---this
corresponds to the condition $U=E$. Consequently, {\em a resonant
transition is one in which the top atom in a well moves
horizontally to the top of a nearest-neighbor well}; motions
either upwards or downwards are non-resonant.} \label{fig1a}
\end{figure}

This paper focuses on one such ``extreme'' parameter regime. Let
$w$ be the amplitude for an atom to tunnel between neighboring
minima of the standing laser wave, and $U$ be the repulsive
interaction energy between two atoms in the same potential well.
When $w$ is smaller than a value of order $U$, the ground state is
a Mott insulator for certain values of the atomic density or
chemical potential. In this state, the average number of atoms in
each potential well must be an integer, $n_0$ (see
Fig~\ref{fig1a}).
Now consider ``tilting'' this Mott insulator \cite{bloch} {\em
i.e.} placing it under an external potential which decreases
linearly along a particular direction in space. Conceptually, it
is useful to imagine that the atoms carry a fictitious `charge',
and then this potential gradient corresponds to applying a uniform
`electric' field, $E$ (in practice this field is applied by
changing the position of the center of the atomic trap
\cite{bloch}). We measure $E$ in units of energy, defining $E$ to
be the maximal drop in potential energy of an atom moving between
nearest-neighbor minima of the periodic potential (the potential
energy drop depends upon the choice of the nearest neighbor, and
we choose the direction(s) along which the drop is the largest to
define $E$). In almost all Mott insulators consisting of electrons
or Cooper pairs, all reasonable electric fields that can be
achieved in the laboratory are small enough so that the relation
$E \ll w, U$ is well satisfied. Remarkably, in the new atomic
systems significantly larger `electric' fields are easily
achievable: this paper shall discuss the regime $E \sim U$ which
has been explored in the recent experiments of Greiner {\em et
al.} \cite{bloch}. More precisely, we shall discuss the regime
\begin{equation}
|U-E|, w \ll E, U, \label{conds}
\end{equation}
while allowing the ratio $(U-E)/w$ to be arbitrary.

We mention, in passing, another experimental system which has been
studied under conditions analogous to (\ref{conds}). Electron
transport has been investigated in arrays of GaAs quantum dots
\cite{marcus}, when the voltage drop between neighboring quantum
dots (the analog of $E$) is at or above the charging energy
required to make the transition (the analog of $U$). However, in
these systems the excess electron energy can be dissipated away to
the underlying lattice, and so it appears that the threshold
behavior can be described by dissipative classical models
\cite{wingreen}. In contrast, for the atomic systems of interest
in the present paper, there is essentially no dissipation over the
time scales of interest, and a fully quantum treatment must be
undertaken.

It useful to explicitly state our model Hamiltonian for the Mott
insulator for our subsequent discussion. We will consider only
Mott insulators of bosons, although the extension to fermionic
Mott insulators is possible \cite{fermion}. We label the minima of
the periodic potential by lattice sites, $i$, and assume that all
bosons occupy a single band of ``tight-binding'' orbitals centered
on these sites. Let $b_i^{\dagger}$ be the creation operator for a
boson on site $i$. We will study the boson Hubbard model
\cite{mpaf,zoller,monien}
\begin{eqnarray}
\mathcal{H} = -w \sum_{\langle ij \rangle} \left( b_{i}^{\dagger}
b_j+ b_{j}^{\dagger} b_i \right) &+& \frac{U}{2} \sum_i n_i (n_i
-1) \nonumber \\ &-& E \sum_{i} {\bf e} \cdot {\bf r}_i n_i
\label{hubbard}
\end{eqnarray}
where $\langle ij \rangle$ represents pairs of nearest neighbor
sites,
\begin{equation}
n_i \equiv b_i^{\dagger} b_i , \label{ni}
\end{equation}
${\bf r}_i$ are the spatial co-ordinates of the lattice sites (the
lattice spacing is unity), and ${\bf e}$ is a vector in the
direction of the applied electric field (${\bf e}$ is not
necessarily a unit vector---its length is determined by the
strength of the electric field, the lattice structure, and our
definition of $E$ above). We will mainly consider simple cubic
lattices, with the ${\bf e}$ oriented along one of the lattice
directions and of unit length. Not shown in (\ref{hubbard}) is an
implied chemical potential term which is chosen so that the
average density of atoms per site is $n_0$. We will restrict our
attention to the case where $n_0$ is of order unity.

Some simple key points can be made by first considering the
non-interacting case, $U=0$, and also by simplifying to one
spatial dimension\cite{meystre}. For this special case, we can
write $\mathcal{H}$ as
\begin{equation}
\mathcal{H}_0 = -\sum_{\ell} \left( w b_{\ell}^{\dagger} b_{\ell
+1} + w b_{\ell + 1}^{\dagger} b_{\ell} + E \ell
b_{\ell}^{\dagger} b_{\ell} \right) \label{oned}
\end{equation}
where $\ell$ is an integer labelling the lattice sites. The exact
single-particle eigenstates of $\mathcal{H}_0$ can be easily
obtained: the eigenenergies form a Wannier-Stark ladder, and the
most important property of the wavefunctions is that they are all
{\em localized}. Specifically, the eigenstates can be labeled by
an integer $m$ which runs from $-\infty$ to $\infty$, the exact
eigenenergies are
\begin{equation}
\epsilon_m = E m, \label{bloch}
\end{equation}
and the corresponding exact and normalized wavefunctions can be
expressed in terms of Bessel functions:
\begin{equation}
\psi_m (\ell) = J_{\ell-m} (2w/E); \label{bessel}
\end{equation}
for a derivation see {\em e.g.\/} Ref.~\onlinecite{wilkins} (their
analysis is in a different gauge). The $m$'th state is localized
near the site $\ell=m$, and for large $|\ell-m|$ its wavefunction
decays as
\begin{equation}
|\psi_m (\ell )| \sim \exp \left[ - |\ell -m| \ln \left(
\frac{|\ell - m| E}{e w} \right) \right];
\end{equation}
the decay is faster than exponential, and is extremely rapid under
the conditions (\ref{conds}). The reader should resist the
temptation to imagine that a particle placed initially at the site
$\ell$ will eventually be accelerated by the applied electric
field out to infinity. Instead, the particle remains localized
near its initial site, and undergoes {\em Bloch oscillations} with
period $ h/E$; indeed, as is clear from the simple form of
(\ref{bloch}), its wavefunction is exactly equal to its initial
wavefunction at regular time intervals of $h/E$. The particle can
escape to infinity only be a process of Zener tunneling to higher
bands not included in the single band tight-binding models in
(\ref{oned}) and (\ref{hubbard}); the probability of such
tunneling is negligibly small in the experiments of interest here,
and so will be ignored in our analysis.

\begin{figure}
\centerline{\includegraphics[width=2.2in]{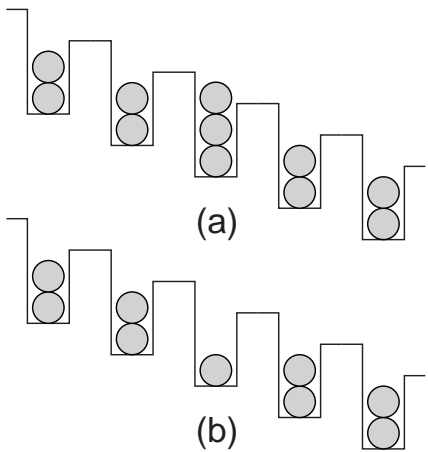}}
\caption{Notation as in Fig.~\ref{fig1a}. ({\em a}) A
quasiparticle on site 3; the motion of this quasiparticle is
described by the localized wavefunctions (\protect\ref{bessel})
but with $w$ replaced by $3w$. ({\em b}) A quasihole on site 3;
the motion of this quasihole is also described by the localized
wavefunctions (\protect\ref{bessel}) but with $w$ replaced by
$2w$.} \label{fig1bc}
\end{figure}

We now return our discussion to the full Hubbard model
(\ref{hubbard}). As was the case in (\ref{bloch}), the spectrum of
this Hamiltonian is unbounded from below for $E \neq 0$, and so it
does not make sense to ask for its ``ground state'' for any
density of particles. Rather, guided by the experimental situation
of Ref.~\onlinecite{bloch}, we are interested in states which are
accessible from the translationally invariant Mott state (with an
average of $n_0$ particles on every site) over the experimentally
relevant time scales. The experiment\cite{bloch} begins at $E=0$
with a Mott insulator with $n_0$ particles per site, rapidly ramps
up $E$ to a value of order $U$, and detects the change in the
state. For $w \ll U$, and for most values of $E$, the experiments
displayed little detectable change in the state of the system. We
can initially understand this by a simple extension of the
argument presented above for the non-interacting model
$\mathcal{H}_0$. Consider a `quasiparticle' state of the Mott
insulator, created by adding a single additional particle on one
site, as shown in Fig~\ref{fig1bc}a.
To leading order in $w/U$, the motion of this quasiparticle along
the direction ${\bf e}$ is described by an effective Hamiltonian
which is identical in form to $\mathcal{H}_0$, but with the
hopping matrix element $w$ replaced $w (n_0 + 1)$. So any such
quasiparticle states created above the Mott insulator will remain
localized and will not have the chance to extend across the system
to create large changes in the initial state. A similar
localization argument applies to the quasihole state shown in
Fig~\ref{fig1bc}b: it experiences an electric force in the
opposite direction, the effective hopping matrix element is $w
n_0$, and all quasihole states are also all localized in the
direction ${\bf e}$. Indeed, it is not difficult to see that the
same localization argument applies to all deformations of the Mott
insulator which carry a net charge.

\begin{figure}
\centerline{\includegraphics[width=2.2in]{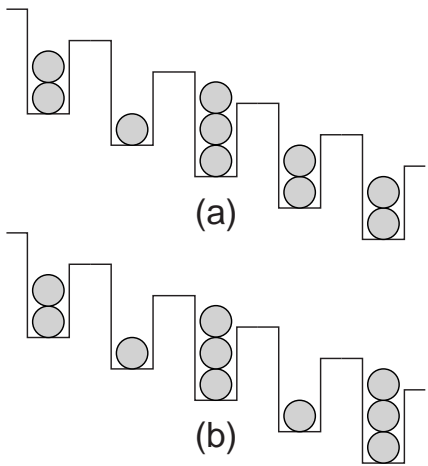}}
\caption{Notation as in Fig.~\ref{fig1a}. ({\em a}) A dipole on
sites 2 and 3; this state is resonantly coupled by an
infinitesimal $w$ to the Mott insulator in ({\em a}) when $E=U$.
({\em b}) Two dipoles between sites 2 and 3 and between 4 and 5;
this state is connected via multiple resonant transitions to the
Mott insulator for $E=U$.} \label{fig1df}
\end{figure}

The important exceptions to the above argument for the stability
of the Mott state are deformations which carry no net charge. It
is the primary purpose of this paper to describe the collective
properties of such neutral states. They will be shown to yield a
resonantly strong effect on the Mott state when $E \sim U$, which
has been dramatically observed in the experiments of Greiner {\em
et al.} \cite{bloch}. Indeed, Greiner {\em et al.} have already
identified an important neutral deformation of the Mott state---it
is the {\em dipole} state consisting of a quasiparticle-quasihole
pair on nearest neighbor sites, as shown in Fig~\ref{fig1df}a.
A key consequence of our discussion above is that, for $w \ll E$
(a condition we assume throughout), we can safely neglect the
independent motion of the quasiparticle and of the quasihole along
the direction of ${\bf e}$. Only their paired motion as dipoles
will be important along ${\bf e}$, although they can move
independently along directions orthogonal to ${\bf e}$.

For $w=0$, the dipole state in Fig~\ref{fig1df}a differs in energy
from the Mott state by $U-E$. So these states become degenerate at
$U=E$, and an infinitesimal $w$ leads to a {\em resonant} coupling
between them. However, there are a large number of other states
which are resonantly coupled to one of more of these states, and
they also have to be treated on an equal footing. Indeed, it is
sufficient for an given state to be resonantly coupled to any one
other state in the manifold of resonant states for it to be an
equal member of the resonant family---it is {\em not\/} necessary
to have a direct resonant coupling to the parent Mott insulator.
The reader should already notice that multiple dipole deformations
of the Mott insulator (like the state in Fig~\ref{fig1df}b) are
part of the resonant family. In dimensions greater than one, these
dipole states are only a small fraction of the set of resonant
states, as we will show below. We are now in a position to
succinctly state the purpose of this paper: {\em identify the
complete set of states resonantly coupled to the Mott state under
the conditions (\ref{conds}), obtain the effective Hamiltonian
within the subspace of these states, and determine its spectrum
and correlations.} The results will allow us to address the strong
response of the Mott insulator to an electric field $E \sim U$
observed by Greiner {\em et al.}\cite{bloch}, and lead to some
definite predictions which can be tested in future experiments.

\begin{figure}
\centerline{\includegraphics[width=2.2in]{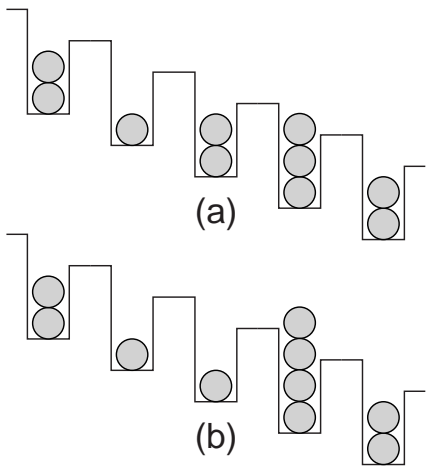}}
\caption{Notation as in Fig.~\ref{fig1a}. Two states which are not
part of the resonant manifold. ({\em a}) An attempt to create
dipoles between sites 2 and 3 and also between sites 3 and 4; the
result is a single dipole of length 2 which has energy $U-2E$
relative to the Mott insulator, and so this long dipole is {\em
not} part of the resonant family of states. ({\em b}) A state with
energy $3(U-E)$ relative to the Mott insulator; this state is not
part of the resonant family because its largest effective matrix
element to any state in the resonant family is of order $w^2/U$
(for $U=E$; see (\protect\ref{w2})). In contrast, all states
within the resonant family are connected to at least one other
state also in the family by a matrix element of order $w$.}
\label{fig1eg}
\end{figure}

The first step in our program is a complete description of the set
of resonant states. We will do this first for one dimension in the
Section~\ref{sec:oned}, and for all higher dimensions in
Section~\ref{sec:3d}. The effective Hamiltonian in the resonant
subspace will be shown to contain strong correlations among its
degrees of freedom, but we will demonstrate that these can be
satisfactorily treated by available analytic and numerical methods
in many body theory. Before embarking on a detailed description of
our computation, the reader may find it useful to examine
Figs~\ref{fig1df} and~\ref{fig1eg} for an understanding of the
origin of the strong correlations in the one dimensional case.
Fig~\ref{fig1df} contains only dipole states: notice that while
resonant dipole states can be created separately on nearest
neighbor links, it is not possible to create two dipoles
simultaneously on such links (as in Fig~\ref{fig1eg}a) without
violating the resonant conditions. This implies an infinite
repulsive interaction between nearest neighbor dipoles in the
effective Hamiltonian. Two (or more) dipoles can be safely created
when they are further apart, as shown in Fig~\ref{fig1df}b. Thus
the dipole resonances are not independent of each other, and the
wavefunction contains non-trivial `entanglements' between them.

\subsection{One dimension}
\label{sec:oned}

It is not difficult to see that, in one spatial dimension, the set
of all nearest-neighbor dipole states constitute the entire family
of states resonantly coupled to the Mott insulator in
Fig~\ref{fig1a} for $U=E$ and an infinitesimal $w$. The only
subtlety concerns states like those in Fig~\ref{fig1eg}b, which
are not made up of nearest-neighbor dipoles. For $w=0$, this state
has energy $3(U-E)$ relative to that in Fig~\ref{fig1a}. However
reaching the state in Fig~\ref{fig1eg}b from any state in the
resonant family requires a detour through a non-resonant state. A
simple second-order perturbation theory calculation shows that the
closest state from the resonant family connected to
Fig~\ref{fig1eg}b is a state with dipole between sites 3 and 4,
and that the effective matrix element between them is
\begin{equation}
\frac{w^2n_0 \sqrt{n_0 (n_0 + 1)}}{2} \left( \frac{1}{U} +
\frac{1}{2U-E} \right); \label{w2}
\end{equation}
this is negligibly small, under the conditions (\ref{conds}),
compared to the non-zero matrix elements ($=w$) between states
within the resonant family. Hence we can safely neglect the state
in Fig~\ref{fig1eg}b. More completely, the argument is that after
we diagonalize the Hamiltonian within the resonant family, states
coupled to that in Fig~\ref{fig1eg}b will differ from it by an
energy of order $w$; the coupling in (\ref{w2}) will then be too
weak to induce a resonance.

It is convenient now to introduce bosonic dipole creation
operators, $d^{\dagger}_{\ell}$, to allow us to specify the
resonant subspace and its effective Hamiltonian. Let $|{\rm M} n_0
\rangle$ be the Mott insulator with $n_0$ particles on every site
(the state in Fig~\ref{fig1a} is $|{\rm M} 2 \rangle$). We
identify this state with the dipole vacuum $|0 \rangle$. Then the
single dipole state is
\begin{equation}
d^{\dagger}_{\ell} |0 \rangle \equiv \frac{1}{\sqrt{n_0 (n_0 +
1)}} b_{\ell} b^{\dagger}_{\ell+1} |{\rm M} n_0 \rangle
\label{ddef}
\end{equation}
Notice that we have placed the dipole operator on the left edge of
the dipole which actually resides on links between the lattice
sites. Clearly, we cannot create more than one dipole resonantly
on the same link: hence the dipoles satisfy an on-site hard core
constraint
\begin{equation}
d_{\ell}^{\dagger} d_{\ell} \leq 1. \label{hc1}
\end{equation}
Moreover, we cannot create two dipoles simultaneously on nearest
neighbor links---this leads to a non-resonant state like that in
Fig~\ref{fig1eg}a; such states are prohibited by a hard core
repulsion between nearest neighbor sites
\begin{equation}
d_{\ell}^{\dagger} d_{\ell} d_{\ell+1}^{\dagger} d_{\ell+1} = 0.
\label{hc2}
\end{equation}
The resonant family of states can now be completely specified as
the set of all states of the boson $d_{\ell}$ which satisfy
(\ref{hc1}) and (\ref{hc2}). A typical state is sketched below in
Fig~\ref{fig2}a. Notice that the dipole vacuum, $|{\rm M} n_0
\rangle$ is one of the allowed states.

It is now a simple matter to write down the effective Hamiltonian,
$\mathcal{H}_d$ for the $d_{\ell}$. It costs energy $U-E$ to
create each dipole, and each dipole can be created or annihilated
with an amplitude of order $w$ (this corresponds to the horizontal
motion of particles in Figs~\ref{fig1a}-\ref{fig1eg}). So we have
\begin{equation}
\mathcal{H}_d = - w \sqrt{n_0 (n_0 +1)} \sum_{\ell} \left( d_\ell
+ d_\ell^{\dagger} \right) + (U-E) \sum_\ell d^{\dagger}_\ell
d_\ell \label{hd}.
\end{equation}
The Hamiltonian (\ref{hd}), along with the constraints
(\ref{hc1},\ref{hc2}), constitute one of the correlated many-body
problems we shall analyze in this paper. The eigenstates of
$\mathcal{H}_d$ are characterized by $n_0$ and the single
dimensionless number
\begin{equation}
\lambda \equiv \frac{U-E}{w}, \label{deflambda}
\end{equation}
and a description of their properties as $\lambda$ ranges over all
real values is in Section~\ref{diponed}. (Strictly speaking, the
eigenstates of  $\mathcal{H}_d$ depend only $\lambda/\sqrt{n_0
(n_0 + 1)}$, but $\lambda$ and $n_0$ do not combine into a single
constant in higher dimensions.)

It is interesting to note that there is no explicit hopping term
for the $d_{\ell}$ bosons in $\mathcal{H}_d$: it appears that the
bosons only only allowed to be created from, and to disappear
into, the vacuum by the first term in (\ref{hd}). However, this is
misleading: as we will see in Section~\ref{diponed}, the
combination of the terms in (\ref{hd}) and the constraint
(\ref{hc2}) does generate a local hopping term for the $d_{\ell}$
bosons (see (\ref{hdeff})). Additional dipole hopping terms also
arise from virtual processes of order $w^2/U$ in the underlying
Hubbard model $\mathcal{H}$; however, these are negligibly small
compared to those just mentioned and do not need to be included in
$\mathcal{H}_d$.

We close this subsection by noting that the Hamiltonian
$\mathcal{H}_d$ in (\protect\ref{hd}) and the constraints
(\protect\ref{hc1},\protect\ref{hc2}) can also be written in the
form of a quantum spin chain. We identify the dipole
present/absent configuration on a site $\ell$ as a pseudospin
$\sigma^z_{\ell}$ up/down ($\sigma^{x,y,z}$ are the Pauli
matrices). Then $\sigma^z_{\ell} = 2 d^{\dagger}_{\ell} d_{\ell}
-1$ and
\begin{eqnarray}
\mathcal{H}_d &=& \sum_{\ell} \Bigl[ -w\sqrt{n_0 (n_0 + 1)}
\sigma^{x}_{\ell} + (U-E)(\sigma^z_{\ell} + 1)/2 \nonumber \\
&~&~~~~~~~~~+ J (\sigma^z_{\ell} + 1)(\sigma^z_{\ell+1} + 1)
\Bigr].
\end{eqnarray}
The constraint (\protect\ref{hc2}) is implemented by taking the $J
\rightarrow \infty$ limit of the last term. The spin chain model
so obtained is an $S=1/2$ Ising spin chain in {\em both}
transverse and longitudinal fields. This is known not to be
integrable for finite $J$, but it does appear that the problem
simplifies in the $J \rightarrow \infty$ limit we consider here.

\subsection{Higher dimensions}
\label{sec:3d}

We consider here only hypercubic lattices in $D$ spatial
dimensions, with ${\bf e}$ oriented along one of the principal
cubic axes and a lattice spacing in length ({\em e.g.\/} $D=3$ and
${\bf e} = (1,0,0)$). Other lattices, and other directions of
${\bf e}$, also allow for interesting correlated phases and these
will be mentioned in Section~\ref{other}.

\begin{figure}
\centerline{\includegraphics[width=3in]{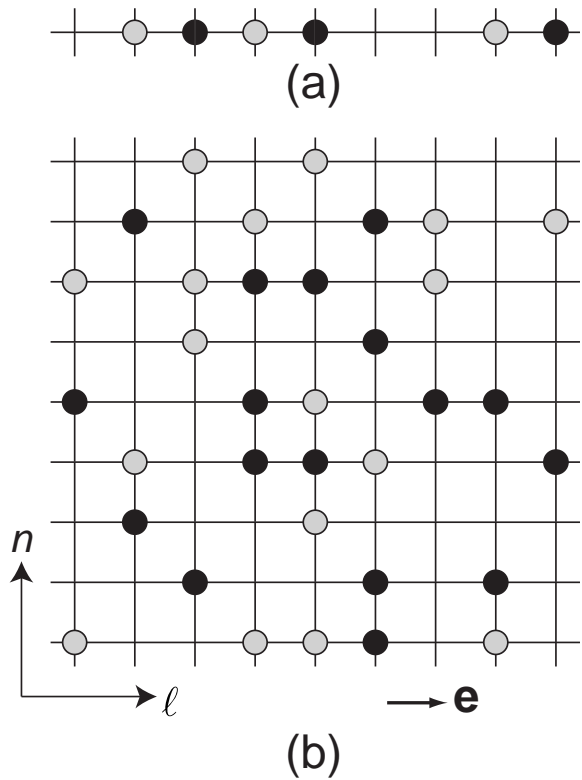}}
\caption{Typical states in the resonant subspace for ({\em a\/})
$D=1$ and ({\em b\/}) $D=2$. Black circles represent sites with
quasiparticles (these sites have $n_{i} = n_0+1$ (see
(\protect\ref{ni})), grey circles represent quasiholes (these
sites have $n_{i}=n_0 -1$), while the remaining sites have
$n_{i}=n_0$. Note that $Q_{\ell}$ in (\protect\ref{hc4}) is zero
for each column {\em i.e.} the total number of quasiparticles in
every column equals the total number of quasiholes in the column
to its immediate left. Only in $D=1$ does this constraint imply
that all states contain only nearest-neighbor dipoles.}
\label{fig2}
\end{figure}

Extension of our reasoning above quickly shows that the dipole
states now constitute only a negligibly small fraction of the set
of states in the resonant family. Once a dipole has been created
on a pair of sites separated by the vector ${\bf e}$, its
quasiparticle and quasihole constituents can move freely and
resonantly, with matrix elements of order $w$, in the $(D-1)$
directions orthogonal to ${\bf e}$. Allowing this process to occur
repeatedly (while maintaining some constraints discussed below),
we can build up the set of all resonantly coupled states. A
typical resonant state in $D=2$ is shown in Fig~\ref{fig2}b.
As in Section~\ref{sec:oned}, it is useful to give an operator
definition of the resonant family. To allow us to distinguish
between the directions parallel and orthogonal to ${\bf e}$, we
replace the $D$-dimensional site label $i$, by the composite label
$(\ell, n)$, where $\ell$ is an integer measuring the co-ordinate
along ${\bf e}$ (as in the one-dimensional case), while $n$ is a
label for sites along the $(D-1)$ transverse directions. Rather
than using dipole operators, we now want to work with bosonic
quasiparticle ($p^{\dagger}_{\ell,n}$) and quasihole
($h^{\dagger}_{\ell,n}$) operators, which create states like those
in Fig~\ref{fig1bc}a and Fig~\ref{fig1bc}b respectively. More
precisely, we now identify $|{\rm M} n_0 \rangle$ with
quasiparticle and quasihole vacuum $|0 \rangle$, and so
\begin{eqnarray}
p^{\dagger}_{\ell,n} |0 \rangle &\equiv & \frac{1}{\sqrt{n_0 + 1}}
b_{\ell,n}^{\dagger} |{\rm M} n_0 \rangle \nonumber \\
h^{\dagger}_{\ell,n} |0 \rangle &\equiv & \frac{1}{\sqrt{n_0}}
b_{\ell,n} |{\rm M} n_0 \rangle.
\end{eqnarray}
The set of resonant states can now be specified by a few simple
constraints on these operators, which are the analogs of
(\ref{hc1},\ref{hc2}). First, there are the obvious on-site
hard-core constraints that no site can have more than one particle
or hole:
\begin{eqnarray}
p_{\ell,n}^{\dagger} p_{\ell,n} &\leq & 1 \nonumber \\
h_{\ell,n}^{\dagger} h_{\ell,n} &\leq & 1 \nonumber \\
p_{\ell,n}^{\dagger} p_{\ell,n} h_{\ell,n}^{\dagger} h_{\ell,n} &
= & 0. \label{hc3}
\end{eqnarray}
Additionally, because of the manner in which these quasiparticles
and quasiholes appear from the Mott state, the total number of
quasiparticles in the $D-1$ dimensional layer with co-ordinate
$\ell+1$ must equal the total number of quasiholes in layer
$\ell$:
\begin{equation}
Q_{\ell} \equiv \sum_n \left( p^{\dagger}_{\ell+1,n} p_{\ell+1,n}
- h^{\dagger}_{\ell,n} h_{\ell,n} \right) = 0. \label{hc4}
\end{equation}
While the quasiparticles and quasiholes are allowed to move freely
within each $D-1$ dimensional layer, they cannot move resonantly
out of any layer on their own; this is, of course, related to the
localization of the Wannier-Stark ladder states discussed earlier
in this section.

Continuing the analogy with Section~\ref{sec:oned}, we can now
easily write down the effective Hamiltonian, $\mathcal{H}_{ph}$,
for the quasiparticles and quasiholes which acts on the set of
states defined by (\ref{hc3}) and (\ref{hc4}). The terms in the
first two lines are the same as those already present in
(\ref{hd}), but expressed now in terms of the quasiparticle/hole
operators, while the last line is associated with motion along the
transverse $D-1$ directions:
\begin{eqnarray}
\mathcal{H}_{ph} &=&  - w \sqrt{n_0 (n_0 + 1)} \sum_{\ell,n}
\left( p_{\ell+1,n} h_{\ell,n} +  p_{\ell+1,n}^{\dagger}
h_{\ell,n}^{\dagger} \right) \nonumber \\  &+& \frac{(U-E)}{2}
\sum_{\ell,n}
\left(p_{\ell,n}^{\dagger} p_{\ell,n} + h_{\ell,n}^{\dagger} h_{\ell,n} \right) \label{hph} \\
&-& w  \sum_{\ell,\langle nm \rangle} \left( n_0
h_{\ell,n}^{\dagger} h_{\ell,m} + (n_0 + 1) p_{\ell,n}^{\dagger}
p_{\ell,m} + \mbox{h.c.} \right). \nonumber
\end{eqnarray}
Here $\langle nm \rangle$ represents a nearest neighbor pair of
sites within a single $(D-1)$ dimensional layer orthogonal to
${\bf e}$. Notice that all the $Q_{\ell}$ in (\ref{hc4}) commute
with $\mathcal{H}_{ph}$, as is required for the consistency of our
approach. As was the case in one dimension, the properties of
$\mathcal{H}_{ph}$ are determined by the single dimensionless
constant $\lambda$ in (\ref{deflambda}); these will be described
in Section~\ref{qphd}.

It is worth reiterating explicitly here that upon specialization
to the case of $D=1$ (when the indices $n,m$ only have a single
allowed value and the set $\langle nm \rangle$ is empty), the
Hamiltonian $\mathcal{H}_{ph}$ above is exactly equivalent to the
one dimensional dipole model $H_d$ in (\ref{hd}).

We note in passing that in a manner similar to $\mathcal{H}_d$,
$\mathcal{H}_{ph}$ in (\protect\ref{hph}) can also be written as a
$S=1$ spin model, with the empty/qausiparticle/quasihole states on
a site corresponding to spin states with $S_z = 0, 1, -1$.

The outline of the remainder of the paper is as follows. The
properties the $D=1$ model $\mathcal{H}_d$ will be described in
Section~\ref{diponed}, while the $D>1$ model $\mathcal{H}_{ph}$
will be considered in Section~\ref{qphd}. We discuss extensions of
our results to other lattices and field directions in
Section~\ref{other}. Implications of our results for experiments
appear in Section~\ref{conc}. The appendices contain some
technical discussion on the nature of the quantum phase
transitions found in the body of the paper.

\section{Dipole phases in one dimension}
\label{diponed}

This section will describe the spectrum of the one-dimensional
dipole Hamiltonian $\mathcal{H}_d$ in (\ref{hd}), subject to the
constraints (\ref{hc1}) and (\ref{hc2}).

An essential point becomes clear simply by looking at the limiting
cases $\lambda \rightarrow \infty$ and $\lambda \rightarrow
-\infty$ (the coupling $\lambda$ was defined in
(\ref{deflambda})). For $\lambda \rightarrow \infty$ the ground
state of $\mathcal{H}_d$ is the non-degenerate dipole vacuum $|0
\rangle$. In contrast, for $\lambda \rightarrow -\infty$ the
ground state is doubly degenerate, because there are two distinct
states with maximal dipole number: $(\cdots d_{1}^{\dagger}
d_{3}^{\dagger} d_{5}^{\dagger} \cdots)|0 \rangle$ and $(\cdots
d_{2}^{\dagger} d_{4}^{\dagger} d_{6}^{\dagger} \cdots)|0
\rangle$. This immediately suggests the existence of an Ising
quantum critical point at some intermediate value of $\lambda$,
associated with an order parameter which is a density wave of
dipoles of period two lattice spacings. We will shortly present
numerical evidence which strongly supports this conclusion.

Further analytic evidence for an Ising quantum critical point can
be obtained by examining the excitation spectra for the limiting
$\lambda$ regimes, and noting their similarity to those on either
side of the critical point in the quantum Ising chain \cite{book}.

For $\lambda \rightarrow \infty$, the lowest excited states are
the single-dipoles: $|\ell \rangle = d_{\ell}^{\dagger} |0
\rangle$; there are $N$ such states ($N$ is the number of sites),
and, at $\lambda=\infty$, they are all degenerate at energy $U-E$.
The degeneracy is lifted at second order in a perturbation theory
in $1/\lambda$: by a standard approach using canonical
transformations, these corrections can be described by an
effective Hamiltonian, $\mathcal{H}_{d,{\rm eff}}$, that acts
entirely within the subspace of single dipole states. We find
\begin{eqnarray}
&& \mathcal{H}_{d,{\rm eff}} = (U-E) \sum_{\ell} \Bigl[ |\ell
\rangle \langle \ell | \nonumber \\ &&+ \frac{n_0 (n_0 +
1)}{\lambda^2} \left( |\ell\rangle \langle \ell |+  |\ell \rangle
\langle \ell+1 | + | \ell+1 \rangle \langle \ell | \right)\Bigr]
\label{hdeff}
\end{eqnarray}
Notice that, quite remarkably, a local dipole hopping term has
appeared, as we promised earlier at the end of
Section~\ref{sec:oned}. The constraints (\ref{hc1},\ref{hc2})
played a crucial role in the derivation of (\ref{hdeff}). Upon
considering perturbations to $|\ell \rangle$ from the first term
in (\ref{hd}) it initially seems possible to obtain an effective
matrix element between any two states $|\ell \rangle$ and
$|\ell^{\prime} \rangle$. However this connection can generally
happen via two possible intermediate states, $|\ell \rangle
\rightarrow d^{\dagger}_{\ell} d^{\dagger}_{\ell^{\prime}} | 0
\rangle \rightarrow |\ell^{\prime} \rangle$ and $|\ell \rangle
\rightarrow | 0 \rangle \rightarrow |\ell^{\prime} \rangle$, and
the contributions of the two processes exactly cancel each other
for most $\ell$, $\ell^{\prime}$. Only when the constraints
(\ref{hc1},\ref{hc2}) block the first of these processes is a
residual matrix element possible, and these are  shown in
(\ref{hdeff}). It is a simple matter to diagonalize
$\mathcal{H}_{d,{\rm eff}}$ by going to momentum space, and we
find a single band of dipole states. The lowest energy dipole
state has momentum $\pi$: the softening of this state upon
reducing $\lambda$ is then consistent with the appearance of
density wave order of period 2. The higher excited states at large
$\lambda$ consist of multiparticle continua of this band of dipole
states, just as in the Ising chain\cite{book}.

A related analysis can be carried out for $\lambda \rightarrow
-\infty$, and the results are very similar to those for the
ordered state in the quantum Ising chain \cite{book}. The lowest
excited states are single band of domain walls between the two
filled dipole states, and above them are the corresponding
multiparticle continua.

\subsection{Exact diagonalization}
\label{num}

We numerically determined the exact spectrum of $\mathcal{H}_d$
for lattice sizes up to $N=18$. As will be evident below, these
sizes were adequate to reliably extract the limiting behavior of
the $N \rightarrow \infty$ limit.

\begin{figure}
\centerline{\includegraphics[width=3in]{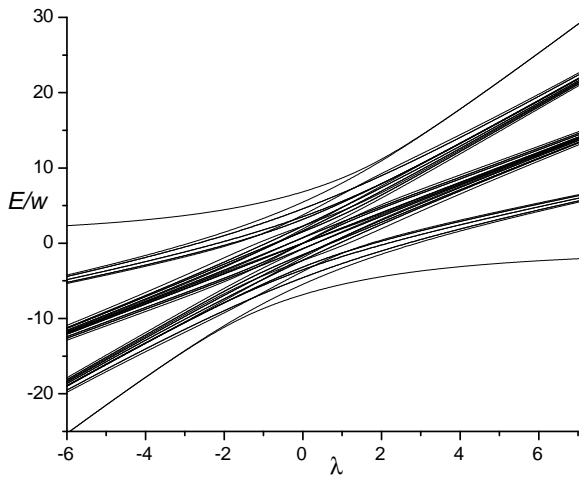}} \caption{All
the eigenvalues of $\mathcal{H}_d$ for $N=8$ and $n_0 = 1$. Note
that the ground state is non-degenerate for positive $\lambda$,
and there are two low-lying levels with an exponentially small
splitting for $\lambda<0$ and $|\lambda|$ large.} \label{levels}
\end{figure}

The complete spectrum of $\mathcal{H}_d$ is shown in
Fig~\ref{levels} for $N=8$ and $n_0 = 1$.
We used periodic boundary conditions on the dipole Hamiltonian in
(\ref{hd}). Note that these do not correspond to periodic boundary
conditions for the original model (\ref{hubbard}); indeed, for
(\ref{hubbard}) the presence of the electric field implies that
periodic boundary conditions are not physically meaningful.
Nevertheless, it is useful to apply periodic boundary conditions
to the translationally invariant effective model (\ref{hd}),
merely as a mathematical tool for rapidly approaching the
$N\rightarrow\infty$ limit. Note that Fig~\ref{levels} shows a
unique ground state for $\lambda \rightarrow \infty$ and a
two-fold degenerate state for $\lambda \rightarrow -\infty$. Above
these lowest energy states, there is a finite energy gap, and the
excited states have clearly split into bands corresponding to the
various ``particle'' continua; these ``particles'' are dipoles for
$\lambda \rightarrow \infty$, and domain-walls between the two
ground states for $ \lambda \rightarrow - \infty$, as we discussed
in the perturbative analysis above.

\begin{figure}
\centerline{\includegraphics[width=3in]{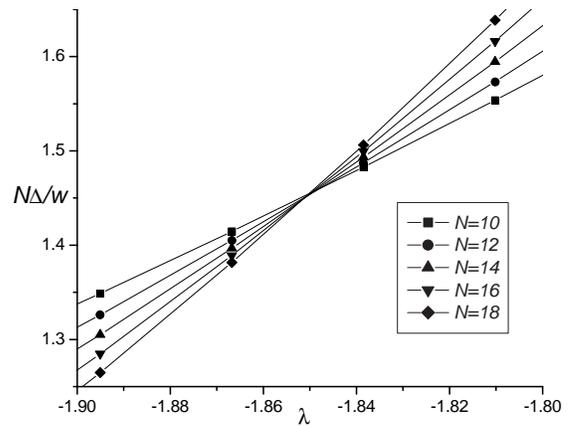}} \caption{The
spacing between the lowest two eigenvalues of $\mathcal{H}_d$
($=\Delta$) as a function $\lambda$ for various system sizes and
$n_0 = 1$. We used periodic boundary conditions for
$\mathcal{H}_d$.} \label{gap}
\end{figure}

We test for a quantum critical point at intermediate values of
$\lambda$ by plotting the energy gap, $\Delta$, in Fig~\ref{gap}.
This gap is the spacing the between the lowest two of the
eigenvalues plotted in Fig~\ref{levels} (for finite system sizes,
these low-lying levels are always non-degenerate). It becomes
exponentially small in the system size as we approach the two
degenerate ground states which are present for $\lambda$
sufficiently negative. In the opposite limit, $\Delta$ approaches
a finite non-zero value, which becomes $U-E$, for $\lambda$ large
and positive. If these two phases are separated by a quantum
critical point, we expect the energy gap to scale as $\Delta \sim
N^{-z}$ at the critical point $\lambda=\lambda_c$, where $z$ is
the dynamic critical exponent. The Ising critical point has $z
=1$, and so Fig~\ref{gap} plots $N \Delta$ as a function of
$\lambda$. We observe a clear crossing point at $\lambda_c \approx
-1.850$ which we identify as the position of the Ising quantum
phase transition. Note that the critical point is shifted away
from the naive value $E=U$ ($\lambda=0$) to $E> U$ because of
quantum fluctuations associated with the hopping of the dipoles.

\begin{figure}
\centerline{\includegraphics[width=3in]{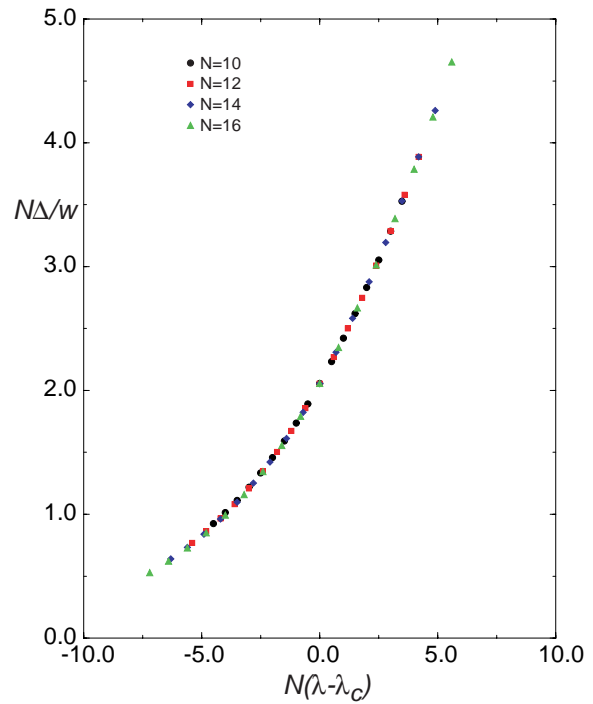}} \caption{Scaling
plot of the energy gap to test for (\protect\ref{scale}). We used
$\lambda_c = -1.850$ and $n_0 = 1$.} \label{nu}
\end{figure}

A second test of Ising criticality is provided by also rescaling
the horizontal axis of Fig~\ref{gap} with $N$. General finite size
scaling arguments imply that the energy gap should obey the
scaling form
\begin{equation}
\Delta = N^{-z} \phi \left(N^{1/\nu} (\lambda-\lambda_c)\right)
\label{scale}
\end{equation}
where $\phi$ is a universal scaling function, and $\nu$ is the
correlation length exponent. We test for (\ref{scale}) in
Fig~\ref{nu} with the Ising exponent $\nu=1$, and again find
excellent agreement.

\begin{figure}
\centerline{\includegraphics[width=3in]{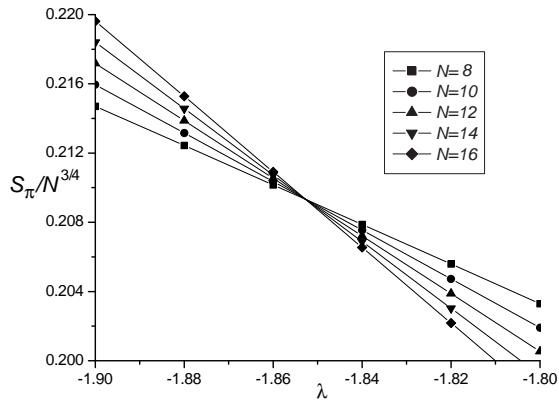}}
\caption{Scaling plot of numerical results for the order parameter
structure factor, $S_\pi$, defined in (\protect\ref{sfacdef}). We
used $n_0 = 1$.} \label{sfac}
\end{figure}

A final, and most sensitive, test for Ising criticality is
provided by a measurement of the anomalous dimension of the order
parameter. The order parameter is the density of dipoles at
momentum $\pi$, and so we computed its equal-time structure factor
\begin{equation}
S_{\pi} = \frac{1}{N} \left\langle \left( \sum_{\ell} (-1)^{\ell}
d_{\ell}^{\dagger} d_{\ell} \right)^2 \right\rangle.
\label{sfacdef}
\end{equation}
Standard scaling arguments imply that this should scale as
$N^{2-z-\eta}$ at $\lambda=\lambda_c$, where $\eta$ is the
anomalous dimension of the order parameter. Using the Ising
exponent $\eta=1/4$, we expect $S_{\pi} \sim N^{3/4}$. This is
tested in Fig~\ref{sfac}.
Note that there is an excellent crossing point at $\lambda_c
\approx -1.853$. This position of the crossing point is completely
consistent with the crossing point found in Fig~\ref{gap}. Thus
Fig~\ref{sfac} provides strong evidence for the expected Ising
exponent $\eta =1/4$. We have also examined a plot which scales
the horizontal axis in Fig~\ref{sfac} as in Fig~\ref{nu}: the data
collapse is again excellent.

\section{Quasiparticle and quasihole phases in higher dimensions}
\label{qphd}

This section will discuss the properties of the $D>1$ dimensional
model of the $p_{\ell,n}$ quasiparticles and $h_{\ell,n}$
quasiholes described the Hamiltonian $\mathcal{H}_{ph}$ in
(\ref{hph}), subject to the constraints (\ref{hc3},\ref{hc4}).

As in Section~\ref{diponed}, it is instructive to first look at
the two distinct limiting values of $\lambda$. The nature of the
ground states is very similar to those in $D=1$ for these ranges
of $\lambda$. For $\lambda \rightarrow \infty$, we have a unique
ground state which contains only small perturbations from the
quasiparticle and quasihole vacuum $|0 \rangle$. For $\lambda
\rightarrow -\infty$, it is clear that we want to maximize the
total number of quasiparticles and quasiholes in the ground state,
subject to the constraints (\ref{hc3},\ref{hc4}). There are a very
large number of ways of doing this, but by  considering
perturbative corrections to the ground state energy in powers of
$1/|\lambda|$, it is not difficult to see that it pays to choose
one of two regular arrangements, in which the occupation numbers
are independent of $n$: $\prod_n \prod_{\ell~{\rm even}}
p^{\dagger}_{\ell+1,n} h^{\dagger}_{\ell,n} |0 \rangle$ or
$\prod_n \prod_{\ell~{\rm odd}} p^{\dagger}_{\ell+1,n}
h^{\dagger}_{\ell,n} |0 \rangle$. So there is a two-fold
degenerate ground state for $\lambda < 0$ and $|\lambda|$ large,
associated with a broken translational symmetry and the
development of density wave order of period 2 in the longitudinal
direction, in both the quasiparticle and quasihole densities.

The excitation spectrum in the limiting ranges of $\lambda$ can
also be determined as in Section~\ref{diponed}. However, the
computations are more involved and we limit ourselves to an
analysis of the $\lambda \rightarrow \infty$ case in
Section~\ref{sc}. We will investigate physics at intermediate
values of $\lambda$ in the subsequent subsections, where we will
see that the possibilities are richer than the appearance of a
single Ising quantum critical point between the states just
discussed: Section~\ref{mft} will present a mean field theory,
while Appendices~\ref{qft} and~\ref{ising} will discuss continuum
quantum field theories which can describe long-wavelength
fluctuations near the phase boundaries.

\subsection{Excitations for $\lambda$ large and positive}
\label{sc}

There is a large manifold of lowest excited states, all of which
have energy $U-E$, in the limit $\lambda \rightarrow \infty$.
These are the states with exactly one $p$ quasiparticle and one
$h$ quasihole, with the particle on the $D-1$ dimensional layer
$\ell+1$ and the hole on the layer $\ell$. We label these states
by
\begin{equation}
|\ell, n,m \rangle \equiv p_{\ell+1,n}^{\dagger}
h_{\ell,m}^{\dagger} |0 \rangle
\end{equation}
We break the degeneracy between these states by considering
corrections in powers of $1/\lambda$. At order $1/\lambda$, the
term in the last line in (\ref{hph}) will allow the quasiparticle
and the quasihole to hop independently in their own layers, but
will not induce any couplings between states with different values
of $\ell$. The latter appear at order $1/\lambda^2$, when as in
(\ref{hdeff}), a nearest neighbor dipole pair can hop
longitudinally between neighboring layers; again, as in $D=1$, the
constraints (\ref{hc3},\ref{hc4}) play a crucial role in
determining these perturbative corrections. These processes are
described the following effective Hamiltonian for the manifold of
excited states with energy $\approx (U-E)$:
\begin{eqnarray}
\mathcal{H}_{ph,{\rm eff}} &=& (U-E) \sum_{\ell} \Biggl[
\sum_{n,m}
|\ell,n,m \rangle \langle \ell,n,m | \nonumber \\
&-& \frac{1}{\lambda} \sum_{\langle n m \rangle, k} \Bigl(n_0
|\ell,n,k \rangle \langle \ell,m,k | \nonumber \\
&~&~~~~~~~~~+ (n_0+1) |\ell,k,n \rangle
\langle \ell,k,m | \Bigr) \nonumber \\
&+& \frac{n_0 (n_0 + 1)}{\lambda^2} \sum_{n} \Bigl( |\ell,n,n
\rangle \langle \ell,n,n| \nonumber \\
&~& \!\!\!\!\!\!\!\!\!\!\!\!\!\!\!\!\!\!\!\!\!\!\!\! + |\ell,n,n
\rangle \langle \ell+1,n,n| + |\ell+1,n,n \rangle \langle
\ell,n,n| \Bigr) \Biggr] \label{hpheff}
\end{eqnarray}
Note that the first summation over $n,m$ is unrestricted and
ranges independently over the two variables, while the second is
over nearest neighbor pairs $\langle nm \rangle$.

The Hamiltonian $\mathcal{H}_{ph, {\rm eff}}$ can be analyzed by
the standard techniques of scattering theory. The terms within the
first two summations in (\ref{hpheff}) lead to a ``two-particle''
continuum of quasiparticle and quasihole states, while the terms
within the last summation allow these states to scatter and
possibly form a dipole bound state. We first form states with
total transverse momentum ${\bf Q}_{\perp}$, and relative
transverse momentum ${\bf q}_{\perp}$ (these momenta are $D-1$
dimensional vectors)
\begin{equation}
\left|\ell, {\bf Q}_{\perp}, {\bf q}_{\perp} \right\rangle =
\frac{1}{N_{\perp}} \sum_{n,n} e^{i {\bf q}_{\perp} \cdot {\bf
r}_n + i  ({\bf Q}_{\perp}-{\bf q}_{\perp}) \cdot {\bf r}_m}
|\ell,n,m \rangle
\end{equation}
where $N_{\perp}$ is the number of sites in each layer, and ${\bf
r}_n$ are the spatial positions of the sites. In this basis of
states $\mathcal{H}_{ph,{\rm eff}}$  Next, we also transform the
single longitudinal co-ordinate, $\ell$, to a `dipole momentum',
$q_{\parallel}$:
\begin{equation}
\left|q_{\parallel},{\bf Q}_{\perp}, {\bf q}_{\perp} \right\rangle
=\frac{1}{N_{\parallel}} \sum_{\ell} e^{i q_{\parallel} \ell}
\left|\ell, {\bf Q}_{\perp}, {\bf q}_{\perp} \right\rangle
\end{equation}
In this basis of states, $\mathcal{H}_{ph,{\rm eff}}$ takes a form
which makes the mapping to standard scattering theory very
explicit. The total transverse momentum, ${\bf Q}_{\perp}$, and
the longitudinal dipole momentum, $q_{\parallel}$, are conserved,
while there is scattering between different values of ${\bf
q}_{\perp}$:
\begin{eqnarray}
&& \mathcal{H}_{ph,{\rm eff}} ({\bf Q}_{\perp}, q_{\parallel}) =
\sum_{{\bf q}_{\perp}} \left[ \varepsilon_p ({\bf q}_{\perp}) +
\varepsilon_h
({\bf Q}_{\perp} - {\bf q}_{\perp}) \right] \nonumber \\
&&~~~~~~~~~~~~~~~~~~~~~~~~~~~ \times \left| q_{\parallel},{\bf
Q}_{\perp}, {\bf q}_{\perp} \right\rangle \left\langle
q_{\parallel},{\bf
Q}_{\perp}, {\bf q}_{\perp}\right| \nonumber \\
&& ~~~~~~~~~+ \frac{w^2 n_0 (n_0 + 1)(1+ 2 \cos
q_{\parallel})}{N_{\perp} (U-E)} \nonumber \\
&& ~~~~~~~~~~~~~~\times \sum_{{\bf q}_{\perp}, {\bf
q}_{\perp}^{\prime}} \left| q_{\parallel},{\bf Q}_{\perp}, {\bf
q}_{\perp} \right\rangle \left\langle q_{\parallel},{\bf
Q}_{\perp}, {\bf q}_{\perp}^{\prime}\right|, \label{scattering}
\end{eqnarray}
where (for a hypercubic lattice)
\begin{equation}
\varepsilon_p ({\bf q}_{\perp}) = \frac{(U-E)}{2} - 2 w (n_0 + 1)
\sum_{\alpha} \cos (q_{\perp \alpha}) \label{pdisp}
\end{equation}
and the summation over $\alpha$ extends over the $D-1$ components
of ${\bf q}_{\perp}$. The expression for $\varepsilon_h ({\bf
q}_{\perp})$ is identical to (\ref{pdisp}) but with $n_0+1$
replaced by $n_0$. The Hamiltonian in (\ref{scattering}) is that
of a particle moving in $(D-1)$ dimensions with momentum ${\bf
q}_{\perp}$ and  dispersion $\varepsilon_p ({\bf q}_{\perp}) +
\varepsilon_h ({\bf Q}_{\perp} - {\bf q}_{\perp})$, scattering off
a delta function potential at the origin with strength $ w^2 n_0
(n_0 + 1)(1+ 2 \cos q_{\parallel})/(U-E)$. Its solution is well
known: in addition to the scattering states, a bound state must be
present in $D-1=1,2$ for any infinitesimal attractive potential,
and for strong enough attraction for $D-1>2$. So for the
physically relevant cases of $D=2,3$, a bound state must form for
a range of $q_{\parallel}$ values near $\pi$. It is clear that the
lowest energy bound state has ${\bf Q}_{\perp} = 0$ and
$q_{\parallel} = \pi$: this is a dipole state, and just as in
$D=1$, it is a precursor to the appearance of longitudinal density
wave order of period 2. The appearance of this dipole bound state
suggests that the first quantum phase transition out of the
featureless and gapped phase present for large positive $\lambda$
is into a state with Ising charge order; however, our discussion
here is for a system with a well-developed gap to quasiparticle
and quasihole states, it is yet not clear whether this approach
continues to hold when the gap becomes small---we will return to
this question in Appendix~\ref{qft}.

\subsection{Mean field theory}
\label{mft}

This section will present the results of a mean field analysis of
$\mathcal{H}_{ph}$. The central idea of the mean field theory is
very simple: we treat the quantum fluctuations along the
longitudinal direction for all $n$ by the exact numerical
treatment developed in Section~\ref{diponed} for $D=1$, while the
transverse couplings are treated in a mean field manner. One
important benefit of this approach is that the important
constraints (\ref{hc3}) are treated exactly.

This approach also naturally suggests the appearance of additional
phases which have no analog in the $D=1$ case. In particular, the
motion of single $p$ and $h$ bosons in the transverse direction
implies that superfluid order can develop along these $D-1$
dimensions only. There is no possibility of superfluidity in the
longitudinal direction because motion along this direction can
occur only via charge neutral dipole pairs which appear in the
first term in (\ref{hph}). This {\em transverse superfluid\/}
therefore has a `smectic' character \cite{kfe}, and its existence
implies that we have to allow for $\langle p \rangle$ and $\langle
h \rangle$ condensates: these appear naturally in our mean field
theory.

As in the mean field treatment of the zero field boson Hubbard
model \cite{mpaf,zoller}, the approximation involves a decoupling
of a hopping term. In particular, we only decouple the last
transverse hopping term in (\ref{hph}), and obtain the following
mean field Hamiltonian for a set of sites, labelled by $\ell$,
representing any chain along the longitudinal direction
\begin{eqnarray}
&& \mathcal{H}_{ph,mf}\left[\langle p_\ell \rangle, \langle
h_{\ell} \rangle \right] = \sum_\ell \Biggl[ -w Z n_0 \left(
\langle h_\ell \rangle
h_\ell^{\dagger}  + \langle h_{\ell} \rangle^{\ast} h_\ell \right) \nonumber \\
&&~~~~~~~~~~~~~~- w Z (n_0 + 1) \left( \langle p_{\ell} \rangle
p_\ell^{\dagger}  + \langle p_{\ell} \rangle^{\ast} p_\ell \right)
\nonumber
\\ &&~~~~~~~~~~~- w \sqrt{n_0 (n_0 + 1)} \left( p_{\ell+1} h_{\ell} +
p^{\dagger}_{\ell+1} h_{\ell}^{\dagger} \right) \nonumber \\
&&~~~~~~~~~~~+  \frac{(U-E)}{2} \left( p_{\ell}^{\dagger} p_{\ell}
+ h_{\ell}^{\dagger} h_{\ell} \right)
 \nonumber \\
&&~~~~~~~~~~~- \mu_{\ell} \left( p_{\ell+1}^{\dagger} p_{\ell+1} -
h_{\ell}^{\dagger} h_{\ell} \right)  \Biggr]. \label{hmf}
\end{eqnarray}
Here $Z$ is the co-ordination number of any site along the $D-1$
transverse directions, and the expectation values $\langle
h_{\ell} \rangle$ and $\langle p_{\ell} \rangle$ have to
determined self-consistently from a diagonalization of (\ref{hmf})
subject to the constraints associated with (\ref{hc3}), which now
become
\begin{equation}
p_{\ell}^{\dagger} p_{\ell} \leq 1~~~;~~~h_{\ell}^{\dagger}
h_{\ell} \leq 1~~~;~~~~p_{\ell}^{\dagger} p_{\ell}
h_{\ell}^{\dagger} h_{\ell} = 0. \label{hcmf}
\end{equation}
We have imposed the constraints (\ref{hc4}) in a mean-field manner
by chemical potentials $\mu_{\ell}$, whose values must be chosen
so that
\begin{equation}
\langle p^{\dagger}_{\ell+1} p_{\ell+1} \rangle = \langle
h_{\ell}^{\dagger} h_{\ell} \rangle \label{qn1}
\end{equation}
is obeyed; note that these constraints are macroscopic, and so
there is no approximation involved in using a chemical potential
to impose them. In practice, the diagonalization of
$\mathcal{H}_{ph,mf}\left[\langle p_\ell \rangle, \langle h_{\ell}
\rangle \right]$ must be carried out for a finite number of sites
$\ell = 1 \ldots N$; we found that the mean-field solutions
approached the $N=\infty$ limit at quite small and manageable
values of $N$. The ground state energy of $\mathcal{H}_{ph}$ per
chain is not equal to the lowest eigenvalue, $E_0$, of
$\mathcal{H}_{ph,mf}\left[\langle p_\ell \rangle, \langle h_{\ell}
\rangle \right]$ but is easily computable from it by the relation
\begin{equation}
E_{ph,mf} = E_0 + w Z \sum_{\ell} \left[ n_0 |\langle h_{\ell}
\rangle|^2 + (n_0 + 1) |\langle p_{\ell} \rangle|^2 \right].
\label{emf}
\end{equation}
Indeed, we can also view $\langle p_{\ell} \rangle$ and $\langle
h_{\ell} \rangle$ as arbitrary complex numbers which are
determined by the minimization of (\ref{emf}).

\begin{figure}
\centerline{\includegraphics[width=3.5in]{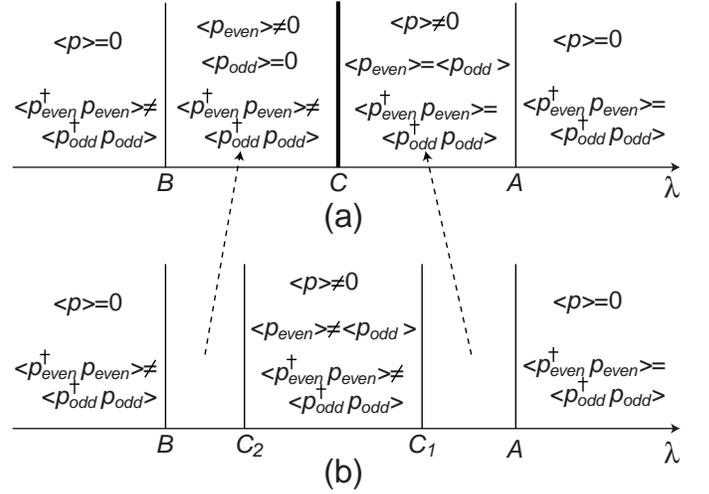}}
\caption{Schematic phase diagrams as a function of $\lambda$. In
({\em a\/}) we display the topology of the phase diagram found by
the solution of the mean field equations: thin lines are second
order quantum phase transitions, while the thick line is a first
order transition. The parity of the $\ell$ index is indicated as a
subscript to the $p$ operators. The expectation values of $h$
quasiholes obey the same relations as those for the $p$
quasiparticles, but with the roles of 'even' and `odd'
interchanged. The Ising density wave order is present for all
$\lambda$ to the left of $C$. In ({\em b\/}) we display a
hypothetical phase diagram, possibly induced by fluctuations, in
which the first order transition is replaced by two second order
transitions; now Ising order is present at $\lambda$ to the left
of $C_1$. There are superfluid-insulator transitions at $A$, $B$,
$C$ and $C_2$, and Ising density wave transitions at $C$ and
$C_1$.} \label{pdiag}
\end{figure}

\begin{figure}
\centerline{\includegraphics[width=3.2in]{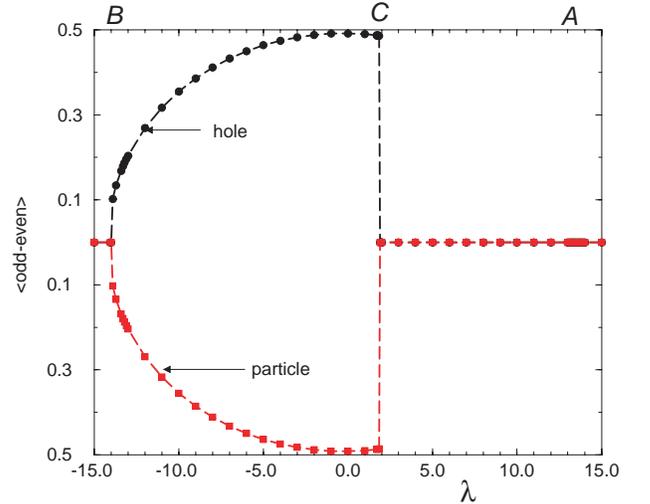}}
\caption{Mean field numerical values of the condensates $\langle
p_{\ell} \rangle$ and $\langle h_{\ell} \rangle$ as a function of
$\lambda$ for $n_0=1$ and $Z=4$. The solutions shown are obtained
by diagonalizing (\protect\ref{hmf}) for $N=6$, but essentially
identical results obtain for $N=4$. } \label{6phase1}
\end{figure}

\begin{figure}
\centerline{\includegraphics[width=3.2in]{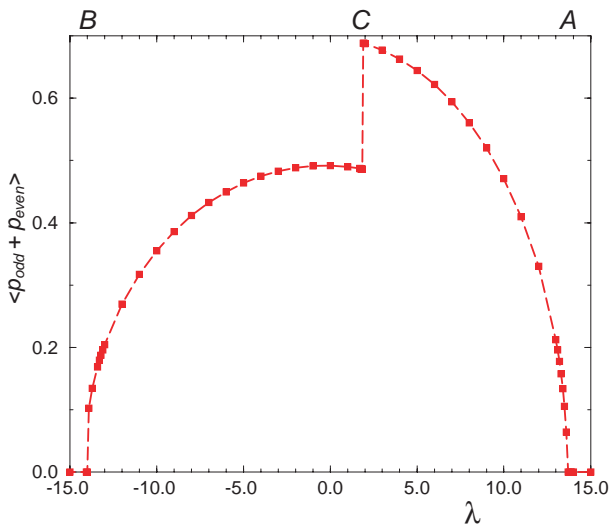}}
\caption{As in Fig~\protect\ref{6phase1}. The values of $\langle
h_{\rm odd}+ h_{\rm even} \rangle$ are very close, but not
identical, to the $p$ values shown above.} \label{6phase2}
\end{figure}

The results of the solution of the above mean-field equations are
summarized in the schematic phase diagrams in Fig~\ref{pdiag} and
in the numerical results in Figs~\ref{6phase1} and~\ref{6phase2}.
It is useful to discuss the phases, in turn, as a function of
decreasing $\lambda$.

For $\lambda$ very large and positive (to the right of the point
$A$ in Fig~\ref{pdiag}), no symmetry is broken, and we have a
featureless ground state with no superfluidity and an energy gap
to all excitations.

There is a superfluid-insulator transition at $A$ driven by the
condensation of the $p$ and $h$ bosons. The superfluidity appears
in the direction transverse to the electric field, and all layers
behave equivalently. We will examine fluctuations near this
critical point in Appendix~\ref{qft} and show that the interlayer
coupling is irrelevant near the critical point in $D=3$, and so
each layer is described by an independent critical theory.

As shown in Fig~\ref{pdiag}a, the mean-field theory exhibits a
first order quantum transition at the the point $C$ associated
with the sudden development of Ising density wave order {\em
i.e.\/} the states with $\lambda$ to the left of $C$ have $\langle
p_{\mbox{even}}^{\dagger} p_{\mbox{even}} \rangle \neq \langle
p_{\mbox{odd}}^{\dagger} p_{\mbox{odd}} \rangle$, and similar for
the density of the $h$ bosons. In mean-field theory, the state to
the immediate left of $C$ also has the loss of the $p$ condensate
in the odd layers (say), and the loss of $h$ condensate in the
even layers. In general, it is quite possible that fluctuations,
beyond those included in the present mean-field theory, will
replace the first order transition at $C$ by two second order
transitions at $C_1$ and $C_2$, as shown in Fig~\ref{pdiag}b. At
the first transition at $C_1$, the order parameter is only the
Ising density wave, while there is $p$ and $h$ transverse
superfluidity in all the layers; the second transition at $C_2$
involves the continuous vanishing of the $p$ ($h$) condensate in
the odd (even) layers in a superfluid-insulator transition, in the
presence of a background of Ising density wave order.

The final transition at the point $B$ involves loss of all $p$ and
$h$ condensates. There is long-range Ising density wave order at
all $\lambda$ to the left of $B$, and a gap to all excitations.

The theory of fluctuations about these mean-field results is
discussed in Appendices~\ref{qft} and~\ref{ising}. As we have
already noted, these could be strong enough to also modify the
topology of the phase diagram in Fig~\ref{pdiag}a. One extreme
possibility is that the transverse superfluid phases could
disappear entirely, and we are left only with two insulating
phases, one with Ising density wave order and the other without;
the phase diagram is then as in $D=1$. However, we show in
Appendix~\ref{qft} that the value of a particular critical
exponent determines that this is not the generic situation.

\section{Other field orientations and lattices}
\label{other}

Our discussion has so far limited itself to hypercubic lattices,
with the direction of the electric field, ${\bf e}$, oriented
along one of the principal axes. Similar analyses can be carried
out for other lattices and for other directions of ${\bf e}$. A
large variety of correlated phases appear possible, including many
not related to those already discussed. We will illustrate these
possibilities by an example here, but leave a more detailed
discussion to future work.

\begin{figure}[t]
\centerline{\includegraphics[width=3in]{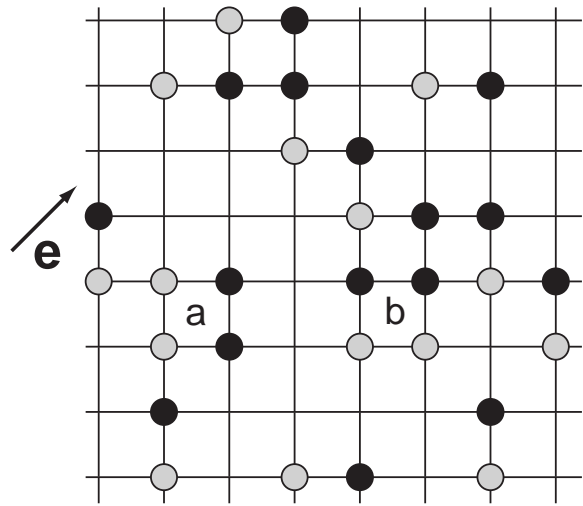}} \caption{A
typical state in the resonant subspace for a square lattice with
${\bf e} = (1,1)$. Representation is as in Fig~\protect\ref{fig2}.
The quasiparticles and quasiholes occur only occur in dipoles
oriented along the $+x$ or $+y$ directions. Note that it is
possible for dipoles to undergo a ring-exchange around a
plaquette, in which the configuration around plaquette $a$ can
become like that around plaquette $b$; this process is contained
in the resonant model $\mathcal{H}_d^{\prime}$ in
(\protect\ref{hdp}) and (\protect\ref{ab}), and does {\em not}
require much weaker virtual processes in the Hubbard model
$\mathcal{H}$ (which are suppressed by powers of $w/U$).}
\label{dimer}
\end{figure}

Consider a square lattice (in $D=2$) but with ${\bf e} = (1,1)$.
In this case, the resonant transitions from the Mott insulator
involve moving a $b_{i}$ boson by one lattice spacing, either
along the $+x$ or the $+y$ direction. However, once such a dipole
has been created, the quasiparticle and the quasihole cannot move
resonantly to any other sites (except by processes of order
$w^2/U$ which we have consistently neglected here). So the
resonant subspace can be described completely in terms of dipole
states, just as in the $D=1$ case discussed earlier. A typical
state is illustrated in Fig~\ref{dimer}.
The effective Hamiltonian of this space of dipole resonant states
is identical in form to (\ref{hd}):
\begin{equation}
\mathcal{H}_d^{\prime} = - w \sqrt{n_0 (n_0 +1)} \sum_{a} \left(
d_a + d_a^{\dagger} \right) + (U-E) \sum_a d^{\dagger}_a d_a
\label{hdp},
\end{equation}
except now the label $a$ extends over the links of the square
lattice. There continues to be a hard-core constraint
$d^{\dagger}_a d_a \leq 1$ like (\ref{hc1}), but the possibility
for new physics arises from the complexity of the generalization
of the constraint (\ref{hc2}), which is now
\begin{equation}
d_{a}^{\dagger} d_a d_b^{\dagger} d_b =0 ~\mbox{for links $a,b$
which share a common site.} \label{ab}
\end{equation}
Note that each dipole blocks the occupancy of dipoles on six
neighboring links. It would be interesting to determine the
properties of $\mathcal{H}_d^{\prime}$ subject to the constraint
(\ref{ab}).

The possibility of rich physics becomes apparent in thinking about
the case $\lambda < 0$ and $|\lambda|$ large. Here the low-lying
manifold of states corresponds to maximizing the number of
dipoles, and these are in one-to-one correspondence with the
close-packed dimer coverings of the square lattice. A natural
ring-exchange term of the dipole bosons also becomes apparent upon
considering perturbative corrections in powers of $1/|\lambda|$:
this derivation is similar in spirit to that in
Section~\ref{diponed} (see Fig~\ref{dimer}). We emphasize that the
dominant ring exchange does not come from virtual higher order
processes in the underlying Hubbard model $\mathcal{H}$ (which are
strongly suppressed by factors of $w/U$), but is already contained
within the physics of the resonant subspace as described by
(\ref{hdp}) and (\ref{ab}). In analogy with other studies of
quantum dimer models \cite{rk,ss,sondhi} and boson ring exchange
models, possibilities of bond-ordered phases open up.
Fractionalized, and Bose metal phases \cite{arun} are also
possible, but these may be more likely on non-bipartite lattices.

We close by noting that it is easily possible to orient ${\bf e}$
so that only one direction is resonant. For a cubic lattice in
$D=3$ this can be done by choosing ${\bf e} = (1,a,b)$ where $a,b
\neq 0,1$ are some arbitrary real numbers. Then resonant
transitions to dipole states can occur only along the $x$
direction, and the resonant manifold separates into decoupled one
dimensional systems, each of which is separately described by the
one-dimensional dipole Hamiltonian $\mathcal{H}_d$ in (\ref{hd}).
This may be a simple way of experimentally realizing the model
$\mathcal{H}_d$.

\section{Implications for experiments}
\label{conc}

An important issue that must be faced at the outset is the extent
to which the non-equilibrium time-dependent experiments can be
described by the ground and low energy states of the effective
models that have been discussed in this paper. In the experiments
of Greiner {\em et al.}\cite{bloch}, the `electric field' (in
practice, this is realized by a magnetic field gradient) is turned
from an initial zero value to $E$ in a time of order $h/w$. In a
system under the conditions (\ref{conds}), this may not allow easy
access of the ground state. As an alternative, we suggest that $E$
be ramped up rapidly to a value to the right of the point A in
Fig~\ref{pdiag}, and then slowly increased through the possible
critical points in Fig~\ref{pdiag}. This could produce states with
either the density wave Ising order, or the transverse superfluid
order.

Having produced such states, the next challenge is to directly
detect the quantum order parameters associated with the phases in
Fig~\ref{pdiag}. We address two possible probes in the subsections
below.

\subsection{Momentum Distribution}
One experimental quantity that is relatively easy to measure is
the momentum distribution of the atoms contained in the optical
lattice.  This is done by shutting off the lattice potential and
the trapping potential and allowing the atoms to freely expand
until the resulting cloud is large enough that its density profile
can be spatially resolved optically.  The scale to which the cloud
expands before measurement can be made much larger than the
original lattice dimensions.  In this limit the final spatial
position at which an atom is detected determines the momentum at
which the momentum distribution function is being measured.

The momentum distribution for the boson Hubbard model containing
$N$ sites is given by
\begin{equation}
\Pi({\bf q}) = |f({\bf q})|^2 \frac{1}{N}\sum_{j,k}e^{i{\bf
q}\cdot ( {\bf r}_j-\vec {\bf r}_k)} \left\langle b^\dagger_j b_k
\right\rangle, \label{eq:momentumdist}
\end{equation}
where $f({\bf q})$ is the form factor for the tight-binding
orbitals associated with the lattice potential, and the momentum
${\bf q} = m{\bf R}/(\hbar t_{ex})$, where ${\bf R}$ is the
distance from the detection position to the center of the trap,
$m$ is the mass of the atoms, and $t_{ex}$ is the time elapsed in
the expansion (this expression ignores the influence of gravity,
but an appropriate modification is straightforward).
The development of off-diagonal long-range order peaks the
momentum distribution at the values of ${\bf q}$ equal to the
reciprocal lattice vectors of the optical lattice potential, and
has been used as an experimental signature of the superfluid
phase. \cite{mark,bloch}

\begin{figure}
\centerline{\includegraphics[width=3.3in]{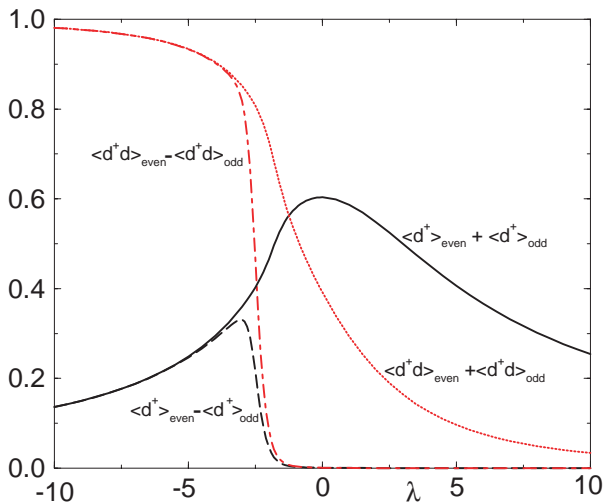}}
\caption{Ground state expectation values of $\langle d_{\ell}
\rangle$ and $\langle d^{\dagger}_{\ell} d_{\ell} \rangle$ for the
$D=1$ model $\mathcal{H}_d$ in (\protect\ref{hd}). The results are
for $N=16$ sites and periodic boundary conditions. A very small
ordering field was applied to choose one of the degenerate Ising
ground states present for sufficiently negative $\lambda$. We have
chosen the gauge in which $\langle d \rangle$ are real.}
\label{expec}
\end{figure}

Let us first consider the $D=1$ case.  A very important
consequence of our restriction to the subspace of resonant states
is that the boson correlator  $\left\langle b^\dagger_{\ell}
b_{\ell'} \right\rangle$ vanishes for $|\ell-\ell'|>1$.  Hence
Eq.~(\ref{eq:momentumdist}) becomes ($q$ is the component of ${\bf
q}$ in the direction of the `electric' field
\begin{eqnarray}
\Pi_{1D}(q) &=& |f(q)|^2 \Biggl[n_0 + \frac{\sqrt{n_0(n_0+1)}}{2
}\nonumber \\
&~&~~~~~~~~\times\sum_{\ell} \left\{ e^{iq}\langle
d^\dagger_{\ell}\rangle + e^{-iq}\langle d_{\ell}\rangle
\right\}\Biggr], \label{eq:momentumdist1D}
\end{eqnarray}
where the lattice spacing has been taken to be unity, and
$d^{\dagger}_{\ell} $ is the dipole creation operator defined in
(\ref{ddef}). For the periodic boundary conditions we have used
(as we noted earlier, such boundary conditions are not physical,
but they should not modify the results in the limit of large
system sizes), the values of $\langle d_{\ell} \rangle$ depend
only on the parity of $\ell$ (a very small ordering field is
applied to lift the Ising symmetry, and choose one of the ground
states in the region with spontaneous Ising order), and hence the
overall amplitude of (\ref{eq:momentumdist1D}) is determined only
by $\langle d_{\rm even} \rangle + \langle d_{\rm odd} \rangle$.
We show our numerical results for these, and other related
quantities, for the Hamiltonian $\mathcal{H}_d$ in (\ref{hd}) in
Figure~\ref{expec}.
There is a broad maximum in $\langle d_{\rm even} \rangle +
\langle d_{\rm odd} \rangle$ near the Ising critical point, as
this is the region with maximal dipole number fluctuations. The
critical singularity in this quantity at $\lambda=\lambda_c$ is
determined by that of the energy operator of the Ising field
theory: this singularity is weak and is essentially unobservable
in Fig~\ref{expec}. The quantities sensitive to the Ising order
parameter (such as $\langle d_{\rm even} \rangle - \langle d_{\rm
odd} \rangle$), show more singular behavior in Fig~\ref{expec}
near $\lambda_c$ determined by the magnetization exponent $\beta$.
However these observables are not detectable by a measurement of
the momentum distribution function.

In higher dimensions ($D>1$) for the case where ${\bf e}$ is
aligned along one of the lattice directions, the dependence of the
distribution function on $q_{\parallel}$ should be qualitatively
similar to the $q$ dependence in the $D=1$ case discussed above.
However a much clearer signal of the transverse superfluidity
should be visible. The presence of the $\langle p \rangle$ and
$\langle h \rangle$ condensates imply that the correlator
(\ref{eq:momentumdist}) has phase coherent contributions when
${\bf r}_j - {\bf r}_k$ lies in the plane perpendicular to the
applied `electric' field. This implies that in states with
transverse superfluidity, there should be Bragg peaks along {\em
lines in ${\bf q}$ space} with values of ${\bf q}_{\perp}$ equal
to the reciprocal lattice vectors of the $D-1$ dimensional lattice
lying in the plane perpendicular to ${\bf e}$. As the transverse
dimensionality is $D-1=2$, the superfluid order can only be
quasi-long-range at nonzero temperature, and hence the Bragg peaks
are not true delta functions in the infinite volume limit, but are
power-law singularities. Experimental detection of these Bragg
lines would be quite interesting.

\subsection{Ising Order Parameter}

We have seen that the Ising order is not directly reflected in the
momentum distribution and hence can not be measured in the free
expansion method described above. The properties of the Ising
order parameter, $\phi$, are discussed in Appendix~\ref{ising};
one convenient definition for $\phi$ is
\begin{equation}
\phi = \frac{1}{N}\sum_\ell (-1)^\ell\left\langle  d^\dagger_\ell
d_\ell\right\rangle
\end{equation}
in $D=1$ (see (\ref{sfacdef}), and a related definition can be
made for higher $D$. One possibility for coupling to the Ising
order parameter experimentally would be to introduce a
phase-locked subharmonic standing wave at half the wave vector of
the optical lattice so that the standing wave takes the form (in
1D)
\begin{equation}
\Phi(x,t) \propto [\cos(Qx)\cos(Qct) +
B\cos(Qx/2+\theta)\cos(Qct/2)].
\end{equation}
Squaring this and taking the time average gives the effective
lattice potential
\begin{equation}
V(x) \propto -[\cos^2(Qx) + B^2\cos^2(Qx/2+\theta)].
\end{equation}
Adjusting the relative phase to $\theta=0$ or $\pi/2$ adds a
`staggered magnetic field' term to the Ising Hamiltonian
\begin{equation}
H_B \propto \pm B^2 \phi.
\end{equation}

A simpler experimental method for the case where the trap
confinement is strong in the directions transverse to the axis of
the 1D lattice is the following.  An additional standing wave
(derived from the same laser) but oriented in the $y$ direction
(say) would yield
\begin{equation}
\Phi(x,y=0,t) \propto [\cos(Qx)\cos(Qct) + B\cos(Qct)]
\end{equation}
and hence a potential along the $x$ axis of
\begin{equation}
V(x,y=0)\propto -[\cos^2(Qx) +2B\cos(Qx) +B^2]
\end{equation}
which would also couple to the Ising order
\begin{equation}
H_B \propto B\phi.
\end{equation}
In either case, such a perturbation could be used to break the
Ising symmetry and selectively populate one of the two Ising
states. In addition, it could be used to {\em measure} the order
parameter itself. The AC stark shift of the atomic hyperfine
levels would differ between adjacent sites. The relative strengths
of the split hyperfine absorption lines would then be a measure of
the Ising order parameter.\cite{markfootnote}

\begin{acknowledgments} We thank Immanuel Bloch and Mark Kasevich
for numerous valuable discussions of their experiments. This
research was supported by US NSF Grants DMR 0098226 and DMR
0196503.
\end{acknowledgments}

\appendix

\section{Fluctuations and quantum field theories: superfluid-insulator transitions}
\label{qft}

The mean field theory of Section~\ref{mft} can be used as a
starting point for a more sophisticated treatment of fluctuations.
Such fluctuations will modify the mean-field exponents in the
vicinity of the second order phase boundaries in Fig~\ref{pdiag}a,
but could also change the topology of the phase diagram to that in
Fig~\ref{pdiag}b.

We analyze fluctuations about the mean field results using a
method very similar to that described in Chapter 10 of
Ref.~\onlinecite{book} for the Hubbard model. We decouple the {\em
intra}-layer hopping terms in $\mathcal{H}_{ph}$ (those in the
last line of (\ref{hph}) only by Hubbard-Stratonovich
transformations using complex fields $P_{\ell} ({\bf r}_{\perp},
\tau)$ and $H_{\ell} ({\bf r}_{\perp}, \tau)$ where ${\bf
r}_{\perp}$ is a spatial co-ordinate for the $D-1$ transverse
directions, and $\tau$ is imaginary time. Then, after standard
simplifications, we obtain an expression for the partition
function $\mathcal{Z}_{ph}$ of $\mathcal{H}_{ph}$ which has the
following schematic form
\begin{eqnarray}
\mathcal{Z}_{ph} &=& \int \mathcal{D} P_{\ell} ({\bf r}_{\perp},
\tau) \mathcal{D} H_{\ell} ({\bf r}_{\perp}, \tau) \nonumber
\\
&~&~~~~~~~~\exp \left[ - \int d^{D-1} {\bf r}_{\perp} \left(
\mathcal{S}_0 + \mathcal{S}_1 \right)\right]. \label{qft1}
\end{eqnarray}
The action $\mathcal{S}_0$ involves couplings only within a single
layer $\ell$, but with different values of ${\bf r}_{\perp}$
\begin{eqnarray}
\mathcal{S}_0 &\equiv& \int d \tau \sum_{\ell} \left[ K_p
|\nabla_{\perp} P_{\ell} ({\bf r}_{\perp}, \tau)|^2 + r_p
|P_{\ell} ({\bf r}_{\perp}, \tau)|^2 \right. \nonumber
\\
&~&~ \left. + K_h |\nabla_{\perp} H_{\ell} ({\bf r}_{\perp},
\tau)|^2 + r_h |H_{\ell} ({\bf r}_{\perp}, \tau)|^2 \right],
\label{qft2}
\end{eqnarray}
and $K_{p,h}$, $r_{p,h}$ are coupling constants. Note that the
factors of $n_0$ and $n_0+1$ in the last line of (\ref{hph}) break
particle-hole symmetry and so there is no special symmetry
relation between these coupling constants. The action
$\mathcal{S}_1$ couples different layers and times together for
the same value of ${\bf r}_{\perp}$
\begin{eqnarray}
e^{- \mathcal{S}_1} &\equiv & \int \mathcal{D} p_{\ell} (\tau)
\mathcal{D} h_{\ell} (\tau) \mathcal{P} \left[ p_{\ell} (\tau),
h_{\ell} ({\tau}) \right] \nonumber \\
&\times &\exp \Biggl[ - \int d \tau \Biggl\{ \sum_{\ell}\left(
p^{\dagger}_{\ell} \frac{\partial p_{\ell}}{\partial \tau} +
h^{\dagger}_{\ell}
\frac{\partial h_{\ell}}{\partial \tau} \right) \nonumber \\
&~&+ \mathcal{H}_{ph,mf} \left[ P_{\ell} ({\bf r}_{\perp}, \tau),
H_{\ell} ({\bf r}_{\perp}, \tau)\right] \Biggr\} \Biggr]
\label{qft3}
\end{eqnarray}
with $\mathcal{H}_{ph,mf}$ defined in (\ref{hmf}), and
$\mathcal{P}$ is a projection operator which represents the
constraints (\ref{hcmf}) (these could be imposed formally in the
functional integral by a very strong on-site repulsive interaction
among the $p_{\ell}$ and $h_{\ell}$ bosons). As in
Section~\ref{mft}, we have imposed the constraints (\ref{hc4}) by
time-independent Lagrange multipliers (``chemical potentials'')
$\mu_{\ell}$: as we noted earlier, there is no approximation
involved in neglecting the fluctuations of $\mu_{\ell}$, because
there is only one constraint per layer and there are a macroscopic
number of particles within each layer. The values of $\mu_{\ell}$
are to be determined at the end by the requirements
\begin{equation}
\frac{\partial \ln \mathcal{Z}_{ph}}{\partial \mu_{\ell}} = 0
\label{qft4}
\end{equation}

Further progress in describing the properties of
$\mathcal{Z}_{ph}$ requires some understanding of the structure of
$\mathcal{S}_1$. This was already addressed to some extent in
Section~\ref{mft} where we explored the properties of the
Hamiltonian $\mathcal{H}_{ph,mf}$. However, here we need to
generalize that analysis to the case where its arguments are
time-dependent fields $P_{\ell} ({\bf r}_{\perp}, \tau)$ $H_{\ell}
({\bf r}_{\perp}, \tau)$. This is quite an involved task, but we
will only need some general constraints that are placed on the
structure of $\mathcal{S}_1$ by the principles of gauge
invariance. In particular, associated with the conservation laws
(\ref{hc4}), we observe that $\mathcal{Z}_{ph}$ is invariant under
the time and layer-dependent transformations generated by the
arbitrary field $\phi_{\ell} (\tau)$
\begin{eqnarray}
p_{\ell+1} &\rightarrow& p_{\ell+1} e^{i \phi_{\ell} (\tau)}
\nonumber
\\
h_{\ell} &\rightarrow& h_{\ell} e^{-i \phi_{\ell} (\tau)}
\nonumber
\\
P_{\ell+1} &\rightarrow& P_{\ell+1} e^{i \phi_{\ell} (\tau)}
\nonumber
\\
H_{\ell} &\rightarrow& H_{\ell} e^{-i \phi_{\ell} (\tau)}
\nonumber \\
\mu_{\ell} &\rightarrow& \mu_{\ell} + i \frac{\partial
\phi_{\ell}}{\partial \tau} \label{gauge}
\end{eqnarray}
We are interested here only in the case of time-independent
$\mu_{\ell}$, and so this transformation takes $\mu_{\ell}$ into
an unphysical set of values; nevertheless, as we will see shortly,
(\ref{gauge}) is still useful in placing constraints on
$\mathcal{S}_1$ in the physical regime.

First, we address the influence of fluctuations by approaching the
transition involving condensation of $P_{\ell}$, $H_{\ell}$ from
the side of large and positive $\lambda$, {\em i.e.} we increase
$E$ (and decrease $\lambda$) until mean-field theory indicates we
are approaching a phase with transverse superfluidity at the point
$A$ in Fig~\ref{pdiag}. The ground state of $\mathcal{H}_{ph,mf}$
is translationally invariant in this region, and so we can safely
assume that all the coupling constants in $\mathcal{S}_1$ are also
independent of $\ell$. Similarly, we can assume that $\mu_{\ell}$
is also independent of $\ell$. If we were to approach the
condensation of $P_{\ell}$, $H_{\ell}$ from the opposite side of
negative $\lambda$, the ground state of $\mathcal{H}_{ph,mf}$
would have a broken Ising symmetry, and the following analysis
would only need to be modified by allowing all couplings, and
$\mu$, to depend upon the $\ell$ sublattice. We describe the
action $\mathcal{S}_1$ by expanding it in powers of the fields
$P_{\ell}$, $H_{\ell}$, and in their temporal gradients (the ${\bf
r}_{\perp}$ and $\tau$ dependence of these fields is now
implicit); to second order in the fields and to first order in
temporal gradients, the most general terms invariant under
(\ref{gauge}) are
\begin{eqnarray}
\mathcal{S}_1 &=& \sum_{\ell} \int d \tau \Biggl[ \widetilde{K}_p
P_{\ell}^{\ast} \frac{\partial P_{\ell}}{\partial \tau} +
\widetilde{K}_h H_{\ell}^{\ast} \frac{\partial H_{\ell}}{\partial
\tau} \nonumber \\ &+& \widetilde{K}_{ph} \left( P_{\ell+1}
\frac{\partial H_{\ell}}{\partial \tau} +  P_{\ell+1}^{\ast}
\frac{\partial H_{\ell}^{\ast}}{\partial \tau} \right) \label{qft5} \\
&+& \widetilde{r}_p |P_{\ell}|^2 + \widetilde{r}_h |H_{\ell}|^2 +
\widetilde{r}_{ph} \left( P_{\ell+1} H_{\ell} + P_{\ell+1}^{\ast}
H_{\ell}^{\ast} \right) \Biggr] \nonumber
\end{eqnarray}
Consistently requiring invariance of (\ref{qft5}) under the
time-dependent gauge transformations (\ref{gauge}) to the order we
have performed the expansion in $\mathcal{S}_1$ demands additional
constraints on the coupling constants above; these are
\begin{equation}
\widetilde{K}_p = - \frac{\partial \widetilde{r}_p}{\partial \mu}
 ~;~\widetilde{K}_h =  \frac{\partial \widetilde{r}_h}{\partial
\mu}~;~  \widetilde{K}_{ph} =  \frac{\partial
\widetilde{r}_{ph}}{\partial \mu}. \label{qft6}
\end{equation}
There are also a large number of permitted higher order terms in
$\mathcal{S}_1$ which we have not written down explicitly; some of
these will play an important role below.

Armed with the low order terms in the action $\mathcal{S}_0 +
\mathcal{S}_1$ controlling the fluctuations of $P_{\ell}$ and
$H_{\ell}$ we can now use standard techniques to focus on the low
energy excitations. It is natural to diagonalize the quadratic
form displayed in these actions: this will lead to two eigenmodes
with distinct eigenvalues. We focus attention on the lower
eigenmode, while integrating out the higher eigenmode. We identify
the lower eigenmode by the field $\Psi_{\ell}$: this has the
structure
\begin{equation}
\Psi_{\ell} ({\bf r }_{\perp},\tau) = c_p P_{\ell+1} ({\bf
r}_{\perp} , \tau) + c_h H_{\ell}^{\ast} ({\bf r}_{\perp}, \tau)
\label{qft7}
\end{equation}
for some constants $c_{p,h}$. Note that we are performing the same
`rotation' in field space for all ${\bf r}_{\perp}$ and $\tau$
(and hence all frequencies). This ensures that $\Psi_{\ell}$ has a
simple behavior under (\ref{gauge}):
\begin{equation}
\Psi_{\ell} \rightarrow \Psi_{\ell} e^{i \phi_{\ell} (\tau)}.
\label{qft8}
\end{equation}
We integrate out the high energy mode orthogonal to (\ref{qft7}),
and obtain our final effective action now expressed in terms of
$\Psi_{\ell}$:
\begin{eqnarray}
&& \!\!\!\!\!\!\!\!\mathcal{Z}_{ph} = \int \mathcal{D} \Psi_{\ell}
({\bf r}_{\perp}, \tau ) \exp \left[ - \int d^{D-1} {\bf
r}_{\perp} d \tau \sum_{\ell}
\left( \mathcal{L}_0 + \mathcal{L}_1 \right)\right] \nonumber \\
&&\mathcal{L}_0 = \left| \nabla_{\perp} \Psi_{\ell} \right|^2 +
\left| \frac{\partial \Psi_{\ell}}{\partial \tau} \right|^2 +
r_{\psi} |\Psi_{\ell}|^2 + \frac{u}{2} |\Psi_{\ell}|^4 \nonumber
\\
&&~~~~~~~~~~+ v
|\Psi_{\ell}|^2 |\Psi_{\ell+1}|^2 \nonumber \\
&& \mathcal{L}_1 = K_{\psi} \Psi_{\ell}^{\ast} \frac{\partial
\Psi_{\ell}}{\partial \tau}. \label{qft9}
\end{eqnarray}
We have rescaled $\Psi_{\ell}$ and time to obtain unit
coefficients for the first two terms in $\mathcal{L}_0$, and the
${\bf r}_{\perp}$ dependence of $\Psi_{\ell}$ is implicit. We have
also written down a quartic non-linearity within a layer ($u$),
and the simplest coupling between neighboring layers ($v$) which
preserves invariance under (\ref{qft8}); we expect both these
couplings to be positive because of the repulsive interactions
between the microscopic bosonic degrees of freedom. The parameter
$r_{\psi}$ tunes the system across the quantum phase transition at
the point $A$ in Fig~\ref{pdiag} which resides at $r_{\psi} =
r_{\psi c}$; the transition is from the featureless, gapped phase
at large positive $\lambda$ ($r_{\psi}> r_{\psi c}$) to a phase
with superfluidity in the transverse $D-1$ dimensions as $\lambda$
is decreased ($r_{\psi} < r_{\psi c}$); the superfluidity is
associated with the condensation of $\Psi_{\ell}$.

Just as in the derivation of (\ref{qft6}), we can also examine the
consequences of time-dependent gauge transformations in
(\ref{gauge},\ref{qft8}) on (\ref{qft9}). This now leads to the
relationship
\begin{equation}
K_{\psi} = - \frac{\partial r_{\psi}}{\partial \mu}. \label{qft10}
\end{equation}
Combined with (\ref{qft4}), the above result now yields a crucial
result. Close to the quantum critical point, the singular free
energy associated with $\mathcal{Z}_{ph}$ is determined directly
by $r_{\psi}$. For this singular term to obey (\ref{qft4}), we
conclude (as also argued in Chapter 10 of Ref~\onlinecite{book})
that $\partial r_{\psi}/\partial \mu = 0$ at $r_\psi = r_{\psi
c}$; (\ref{qft10}) now implies
\begin{equation}
K_\psi=0~~~\mbox{at $r_{\psi}= r_{\psi c}$}. \label{qft11}
\end{equation}
So we can neglect $\mathcal{L}_1$, and the critical theory is
described entirely by ${\cal L}_0$. Within each layer $\ell$, this
theory has the relativistic invariance of $(D-1)+1$ spacetime
dimensions and dynamic critical exponent $z=1$.

Before turning to an examination of the properties of
(\ref{qft9}), we pause to discuss the modifications required to
describe the onset of transverse superfluidity with increasing
$\lambda$ in the region $\lambda < 0$ at the point $B$ in
Fig~\ref{pdiag} (a similar reasoning can also be applied to the
point $C_2$ in Fig~\ref{pdiag}b). Here, long-range Ising order is
already present in $H_{ph,mf}$ for $\lambda$ sufficiently
negative. We can proceed to a description of the superfluid
transition as above, but as noted earlier, all couplings in
(\ref{qft9}) will acquire an $\ell$ dependence which modulates
with period 2. The tuning parameter $r_{\psi}$ will also be
different for even and odd $\ell$. Consequently only $\Psi_{\ell}$
with $\ell$ even (say) will become critical near the transition,
while $\Psi_{\ell}$ with $\ell$ odd remains non-critical and can
be integrated out. The simplest interlayer coupling between
critical modes is now $|\Psi_{\ell}|^2 |\Psi_{\ell+2}|^2$, but its
co-efficient should be small and is likely to be attractive.

We now return to an examination of (\ref{qft9}) for the case of
$\ell$-independent couplings at the transition with $U-E$ positive
at the point $A$ in Fig~\ref{pdiag}. It remains to examine the
consequences of the interlayer coupling $v$ on the standard theory
of the superfluid-insulator transition. At $v=0$, we have the
standard $\varphi^4$ field theory with O(2) symmetry in
$(D-1)+1=D$ spacetime dimensions. As a first step, we can compute
the scaling dimension of $v$ at its critical point. A standard
power-counting argument shows that
\begin{equation}
\mbox{dim}[v] = \frac{2}{\nu} - D = \frac{\alpha}{\nu}
\label{qft12}
\end{equation}
where $\nu$ and $\alpha$ are the standard correlation length and
``specific heat'' exponents in $D$ spacetime dimensions. In $D=3$,
the O(2) fixed point has \cite{campo} $\alpha = -0.015 < 0$, and
so we conclude that $v$ is formally irrelevant. In $D=2$, the very
weak specific heat singularity at the Kosterlitz Thouless
transition suggests the same conclusion.

A more complete analysis of the influence of $v$ can be obtained
by considering a physical susceptiblity for ordering in the
longitudinal direction. As we have seen in Sections~\ref{diponed}
and~\ref{mft}, the simplest allowed ordering is a density wave of
period two. The tendency to this ordering is measured by the
static susceptibility $\chi_{\pi}$
\begin{eqnarray}
\chi_{\pi} &=& \frac{1}{N_{\parallel}} \sum_{\ell,\ell^{\prime}}
\int d^{D-1} {\bf r}_{\perp} d\tau (-1)^{\ell+\ell^{\prime}}
\nonumber \\&~&~~~~~~~~~~~~~ \left\langle |\Psi_{\ell} ( {\bf
r}_{\perp}, \tau)|^2 |\Psi_{\ell^{\prime}} ({\bf 0}, 0)|^2
\right\rangle. \label{qft13}
\end{eqnarray}
Note that this response function is similar to $S_\pi$ in
(\ref{sfacdef}), but we are considering here a zero frequency
response, while (\ref{sfacdef}) involved an equal time correlator.
We can compute (\ref{qft13}) in powers of $v$, and by a familiar
Dyson-type argument, write it as
\begin{equation}
\chi_{\pi} = \frac{\mathcal{C}}{1 - 2 v {\mathcal C}}
\label{qft14}
\end{equation}
where $\mathcal{C}$ is an `irreducible' correlator within a single
layer (it is irreducible with respect to cutting a $v$ interaction
line):
\begin{equation}
\mathcal{C} = \int d^{D-1} {\bf r}_{\perp} d\tau \left\langle
|\Psi_{\ell} ( {\bf r}_{\perp}, \tau)|^2 |\Psi_{\ell} ({\bf 0},
0)|^2 \right\rangle. \label{qft15}
\end{equation}
The computation leading to (\ref{qft14},\ref{qft15}) is the
field-theoretic analog of the computations which lead to a dipole
bound state induced by the interlayer coupling in the
strong-coupling analysis of Section~\ref{sc}. Ignoring the
influence of $v$ on $\mathcal{C}$, standard scaling arguments
imply that $\mathcal{C}$ has a singular part which behaves as
\begin{equation}
\mathcal{C} \sim |r_{\psi} - r_{\psi c}|^{-\alpha}. \label{qft16}
\end{equation}
If we had $\alpha > 0$, then the denominator in (\ref{qft14})
would vanish at some $r_{\psi} > r_{\psi c}$ for any small $v$,
and $\chi_{\pi}$ then diverges: this would imply the presence of
an Ising density wave transition before the onset of
superfluidity. However $\alpha < 0$ in $D=3$, and so this
condition does not apply. Nevertheless, there is a significant
(albeit finite) enhancement of the specific near the O(2) critical
point in $D=3$, and so the instability in $\chi_{\pi}$ may well
occur for a moderate value of $v$. If so, the mean-field phase
diagram would be modified, and the Ising ordered phase would fully
overlap and extend beyond the region with transverse
superfluidity. Indeed, under suitable conditions, the superfluid
phase could also shrink to zero, and we would then have only a
single Ising transition between two insulating phases.
Alternatively, if the Ising fluctuations are weaker, $\chi_{\pi}$
could diverge somewhere in the superfluid phase to the left of $A$
in Fig~\ref{pdiag}, and then the mean-field phase diagram would be
modified to the structure in Fig~\ref{pdiag}b.

\section{Ising phase transition} \label{ising}

In Appendix~\ref{qft} we completed a description of fluctuations
near all the superfluid-insulator transitions in Fig~\ref{pdiag}.
It remains to describe the second order Ising critical point $C_1$
in Fig~\ref{pdiag}b; this we do in the present appendix.

As usual, we expect the Ising phase transition to be realized by a
quantum field theory of a real scalar field $\phi ({\bf r},
\tau)$, where ${\bf r} = (\ell, {\bf r}_{\perp})$ is a $D$
dimensional spatial co-ordinate. The main subtlety here is that
the Ising transition occurs in a background of transverse
superfluid order, and corrections from superflow fluctuations can
lead to anisotropic singular corrections to the critical theory. A
theory of an Ising order parameter coupled to isotropic superflow
fluctuations has been analyzed by Frey and Balents \cite{balents};
here, we will show that the particular anisotropic nature of both
the superfluid and Ising order leads to a more singular coupling
between the two order parameters.

Any observable sensitive to the period 2 modulation in the density
of particles or holes can be used to define the order parameter
$\phi({\bf r},\tau)$. A convenient choice in our present continuum
formulation is to take
\begin{equation}
\phi (\ell, {\bf r}_{\perp}, \tau) \sim (-1)^{\ell} |\Psi_{\ell}
({\bf r}_{\perp}, \tau) |^2.  \label{qft17}
\end{equation}
An effective action, $\mathcal{S}_\phi$, for the Ising field
$\phi$ can be generated by using $\phi$ as a Hubbard-Stratonovich
field to decouple the $v$ term in (\ref{qft9}). This leads to an
action with the structure
\begin{eqnarray}
\mathcal{S}_{\phi} &=& \int d^D {\bf r} d \tau \Biggl[ \frac{1}{2}
\left( \partial_{\tau} \phi \right)^2 + \frac{K_{\perp}}{2} \left(
\nabla_{\perp} \phi \right)^2 \nonumber
\\ &~&~~~~~~~~~~~+ \frac{K_{\parallel}}{2} \left(
\nabla_{\parallel} \phi \right)^2 + u_I \phi^4 \Biggl] \nonumber
\\ &-& w_I \sum_{\ell} \int d^{D-1} {\bf r}_{\perp} d \tau (-1)^{\ell}
|\Psi_{\ell} ({\bf r}_{\perp} , \tau)|^2 \phi (\ell, {\bf
r}_{\perp}, \tau),\nonumber \\
&~& \label{qft18}
\end{eqnarray}
where the fluctuations of $\Psi_{\ell}$ are described by
(\ref{qft9}), and we have included the usual analytic terms
present in the $\phi^4$ theory of an Ising quantum critical point.
The last term in (\ref{qft18}) represents a linear coupling
between the Ising order parameter and density fluctuations in the
superfluid state. In the isotropic case considered by Frey and
Balents such a linear coupling was absent, and the simplest
allowed coupling was between $\phi^2$ and the density
fluctuations: this was because the Ising order parameter
represented a density wave at a large wavevector, and they coupled
linearly only to fluctuations of the superfluid phase at the same
wavevector, and the latter are quite high energy. In the present
case also, the $(-1)^{\ell}$ factor in the last term in
(\ref{qft18}), also shows that $\phi$ couples linearly to the
superfluid phase fluctuations at a wavevector $q_{\parallel} =
\pi$. However, the key difference here is that the superfluidity
is present only along the transverse direction, and, to leading
order, the superfluid phase fluctuations are {\em independent\/}
of $q_{\parallel}$.

The singular effect of the $w_I$ term in (\ref{qft18}) can be
illustrated by integrating out the $\Psi_{\ell}$ using the action
(\ref{qft9}) in a single loop approximation. To leading order in
$u$, we are in the transverse superfluid state as long as
$r_{\psi} < 0$, and a simple calculation of the phase and
amplitude fluctuations of the superfluid order parameter shows
that we generate the following term in $\mathcal{S}_{\phi}$:
\begin{equation}
\frac{1}{2u} \sum_{q_{\parallel},{\bf q}_{\perp}, \omega} |\phi
(q_{\parallel},{\bf q}_{\perp}, \omega)|^2 \frac{ |r_{\psi}| ({\bf
q}_{\perp}^2 + \omega^2)}{|r_{\psi}| ({\bf q}_{\perp}^2 +
\omega^2)+ K_{\psi}^2 \omega^2/2}, \label{qft19}
\end{equation}
where $\omega$ is an imaginary frequency. Note that this is a
singular function of ${\bf q}_{\perp}$ and $\omega$ only when
$K_{\psi} \neq 0 $. We do not expect $K_{\psi} = 0$ near the Ising
critical point, because exact particle-hole symmetry is not
present in the underlying Hamiltonian, and the arguments which
lead to (\ref{qft11}) hold only at the superfluid-insulator
transition. All the couplings in $\mathcal{S}_{\phi}$ can be
expected to be a smooth function of $\mu$, and the constraint is
now expected to lead only to a Fisher renormalization\cite{fisher}
of exponents. An analysis of $\mathcal{S}_{\phi}$ with
(\ref{qft19}) included requires a renormalization group
computation: this we leave to future work, as a full discussion of
the renormalization of the momentum dependence of the propagator
requires a two loop analysis.

In closing, we note that although $K_{\psi} \neq 0$, in practice
the degree of particle-hole symmetry breaking is quite small, as
is indicated by the almost equal values of $\langle p \rangle$ and
$\langle h \rangle$ in Figs~\ref{6phase1} and~\ref{6phase2}. So
$K_{\psi}$ can also be expected to be quite small, and we should,
therefore, also consider the case $K_{\psi}=0$. In this case,
(\ref{qft19}) does not induce any singular terms, and we have to
consider terms induced by $\Psi_{\ell}$ fluctuations at higher
orders in $u$, and also the term included in
Ref~\onlinecite{balents}.



\begin{thebibliography}{}

\bibitem{mark} C.~Orzel, A.~K.~Tuchman, M.~L.~Fenselau, M.~Yasuda, and
M.~A.~Kasevich, Science {\bf 291}, 2386 (2001).

\bibitem{bloch} M.~Greiner, O.~Mandel, T.~Esslinger, T.~W.~H\"ansch, and
I.~Bloch, Nature {\bf 415}, 39 (2002).

\bibitem{marcus} C.~I.~Duru\"{o}z, R.~M.~Clarke, C.~M.~Marcus, and
J.~S.~Harris, Jr., Phys. Rev. Lett. {\bf 74}, 3237 (1995).

\bibitem{wingreen} A.~A.~Middleton and N.~S.~Wingreen, Phys. Rev.
Lett. {\bf 71}, 3198 (1993).

\bibitem{fermion} If we ignore the spin of the fermions (as may be the
justified under certain physical conditions, such as the presence
of a magnetic field), then the generalization of our results to
fermionic Mott insulators (which must have $n_0 = 1$) is
relatively straightforward. The $D=1$ results apply unchanged to
fermionic Mott insulators. For $D>1$, the Hamiltonian
$\mathcal{H}_{ph}$ in (\protect\ref{hph}) applies unchanged, but
with $p$ and $h$ fermionic operators. The first two constraints in
(\protect\ref{hc3}) are now automatically accounted for by
fermionic statistics, while the last must be implemented by an
infinite local repulsion between quasiparticles and quasiholes.
Gapped phases appear for large $|\lambda|$ (the region with
$\lambda \ll 0$ having Ising density wave order), but the
particle-hole hopping asymmetry allows this fermionic model to
exhibit gapless Fermi surfaces for motion in the transverse
direction. The case of spinful fermions will have an even richer
behavior, driven by the antiferromagnetic coupling between the
spins.

\bibitem{mpaf} M.~P.~A.~Fisher, P.~B.~Weichman, G.~Grinstein, and
D.~S.~Fisher, Phys. Rev. B {\bf 40}, 546 (1989).

\bibitem{zoller} D.~Jaksch, C.~Bruder, J.~I.~Cirac, C.~W.~Gardiner,and
P.~Zoller, Phys. Rev. Lett. {\bf 81}, 3108 (1998).

\bibitem{monien} T.~D.~Kuhner, S.~R.~White, and H.~Monien,
Phys. Rev. B {\bf 61}, 12474 (2000).

\bibitem{meystre} J.~Zapata, A.~M.~Guzm\'an, M.~G.~Moore, and P.
Meystre, \pra {\bf 63}, 023607 (2001).

\bibitem{wilkins} J.~H.~Davies and J.~W.~Wilkins, Phys. Rev. B
{\bf 38}, 1667 (1998).

\bibitem{book} S.~Sachdev, {\it Quantum Phase Transitions},
Cambridge University Press, Cambridge (1999).

\bibitem{kfe} S.~A.~Kivelson, E.~Fradkin, and V.~J.~Emery, Nature {\bf 393}, 550
(1998).

\bibitem{rk} D.~S.~Rokhsar and S.~A.~Kivelson, Phys. Rev. Lett. {\bf
61}, 2376 (1988).

\bibitem{ss} N.~Read and S.~Sachdev, Phys. Rev. B {\bf 42}, 4568 (1990);
S.~Sachdev, Phys. Rev. B. {\bf 45}, 12377 (1992).

\bibitem{sondhi} R.~Moessner and S.~L.~Sondhi, Phys. Rev. Lett.
{\bf 86}, 1881 (2001).

\bibitem{arun} A.~Paramekanti, L.~Balents, and M.~P.~A.~Fisher,
cond-mat/0203171.

\bibitem{markfootnote} We are grateful to
Mark Kasevich for this suggestion.

\bibitem{campo} M.~Campostrini, A.~Pelissetto, P.~Rossi,
E.~Vicari, Phys. Rev. B {\bf 61}, 5905 (2000).

\bibitem{balents} E.~Frey and L.~Balents,
Phys. Rev. B {\bf 55}, 1050 (1997).

\bibitem{fisher} M.~E.~Fisher, Phys. Rev. {\bf 176}, 257 (1968);
M. Krech, {\em Computer Simulation Studies in Condensed Matter
Physics XII\/}, Eds. D.~P.~Landau, S.~P.~Lewis, and
H.~B.~Schuettler, Springer Verlag, Heidelberg (1999),
cond-mat/9903288.


\end{thebibliography}
\end{document}